\documentclass[12pt]{iopart}

\usepackage{amsfonts, amssymb, mathrsfs, times, float, color, bm}
\usepackage{graphicx}
\usepackage{graphicx,epstopdf}
\usepackage{dcolumn}
\usepackage{bm}
\newcommand{\nn}{\nonumber}

\newcommand{\sss}{\scriptscriptstyle}
\begin{document}

\title[Post-Minkowskian motion in Kerr-Newman spacetime]{Post-Minkowskian solution for the small-deflection motion of test particles in Kerr-Newman spacetime}

\author{Bo Yang$^{\dagger}$, ~~Chunhua Jiang$^{\star}$, ~~Wenbin Lin$^{\star,\dagger}$}%
\address{$^{\dagger}$School of Physical Science and Technology, Southwest Jiaotong University, Chengdu, 610031, China}
\address{$^{\star}$School of Mathematics and Physics, University of South China, Hengyang, 421001, China}
\ead{lwb@usc.edu.cn}

%


\vspace{10pt}

\begin{abstract}
We derive the second-order post-Minkowskian solution for the small-deflection motion of test particles in the external field of the Kerr-Newman black hole via an iterative method. The analytical results are exhibited in the coordinate system constituted by the particles' initial velocity unit vector, impact vector, and their cross-product. The achieved formulas explicitly give the dependences of the particles' trajectory and velocity on the time once their initial position and velocity are specified, and can be applied not only to a massive particle, but also to a photon as well.
\end{abstract}


\vspace{1pc}
\noindent{\it Keywords}: post-Minkowskian solution, small-deflection approximation, Kerr-Newman spacetime

\section{introduction}

One primary object of general relativity is to provide the motion laws for the bodies in the strong gravitational fields. When the bodies' mass is much less than that of the gravitational source and their sizes are much smaller than the characteristic length of the gravitational field, the bodies can be regarded as the test particles, which include the massive particles and the massless ones.

The motion can be classified as the bound motion and the unbound one. For the massive particles, people have mainly focused on the bound motion and achieved a number of the analytical solutions for a variety of  spacetimes~\cite{Hagihara1931,BPTeukolsky1972,Mino2003,KonigsdorfferGopakumar2005,HL2008a,Han2008,FujitaHikida2009,HHLSirimachan2010,GrunauKagramanova2011,PQR2011b,KopeikinEK2012,GKL2012,PQRuffini2013,PoissonWill2014,Wuetal2015,FlathmannGrunau2015,HuangWu2016,COSentorun2016,Jing2017}.
Specifically, Hackmann and Xu~\cite{HackmannXu2013} have classified the colatitudinal and radial motion of a charged particle in the Kerr-Newman spacetime, and formulated the solutions to the motion in terms of the elliptic functions dependent on the Mino time~\cite{Mino2003}. On the other hand, the unbound motion of the massless particles in various spacetimes has been explored in-depth, e,g, the light propagation in the gravitational fields of the (static or moving) multiple bodies~\cite{Will1981,RichterMatzner1982a,RichterMatzner1982b,RichterMatzner1983,Klioner1991,KlionerKopeikin1992,Kopeikin1997,KopeikinSchafer1999,KopeikinMashhoon2002,Klioner2003,KlionerPeip2003,PLTeyssandier2004,WSperhake2004,KopeikinKorobkovPolnarev2006,KopeikinMakarov2007,KopeikinFomalont2007,TeyssandierPoncin-Lafitte2008,Poncin-LafitteTeyssandier2008,ZschockeKlioner2011,Teyssandier2012,DengXie2012,Will2014,HBLePoncin-Lafitte2014,Zschocke2015,SoffelHan2015,Deng2015,Zschocke2016,Deng2016,Zschocke2017}, and the black holes~\cite{BodennerWill2003,Edery2006,IyerHansen2009,Klioner2010,Bozza2010,Gibbons2012,ChakrabortySen2015,LinetTeyssandier2016,Barlow2017} as well. Recently, we have obtained an analytical solution for the photon's unbound motion in the Kerr-Newman spacetime~\cite{JiangLin2018}. Since the motion laws of the massive particles are different from that of the photon, it is of theoretical significance to achieve an unified solution for the unbound test particles, which are not limited to the photon. Moreover, GRAVITY collaboration has successfully measured the orbit of the star S2 around the supermassive black hole candidate Sgr A* in the galactic center, with the pericentre being about 1400 Schwarzschild radii~\cite{GRAVITYCollaboration2018}. This implies that the stellar motion in the strong gravitational field has now become detectable with the current technologies. The future Thirty Meter Telescope (TMT)~\cite{TMT} has higher capabilities for an angular resolution and a ultra-deep spectroscopy, and will provide more measurements of the stellar orbits/trajectories in the strong fields. These observations may serve as one fundamental test of general relativity. Therefore, it is also desirable from the applied views to obtain the analytical solutions to the trajectory and velocity of the unbound bodies. 

When the gravitational fields are not extremely strong but beyond the accuracy of the Newtonian theory, people usually employ the post-Newtonian (PN) or post-Minkowskian (PM) approximations to calculate the motion of the test particles in these kinds of fields. Notice that the PN approximations usually assume that the characteristic velocity of the under-investigation system is non-relativistic. But the motion of the relativistic particles including the photon in the weak gravitational fields are classified into the PN approximations in the Weinberg's textbook\,\cite{Weinberg1972}. So we do not distinguish the PN approximations and the PM ones throughout this paper.

In this work, we derive the trajectory and velocity of the small-deflection particles with/without mass in the gravitational field of a Kerr-Newman black hole via an iterative method in the PM approximations. The iterative method can be used for various spacetimes and up to the arbitrary-order PM approximations under the small-deflection condition. The achieved solutions are formulated in terms of the particles' initial position and velocity and the time in the harmonic coordinates, and do not involve any elliptic functions. One aim of this work is to achieve a good precision in solving the geodetic equation without slow and complicated numerical methods. On the other hand, the analytical solutions are usually preferred since they can exhibit the effects of the source's parameters (mass, charge, and angular momentum) on the particles' motion explicitly. Another goal of this work is to provide an unified description for the test particles' small-deflection motion in the Kerr-Newman spacetime, which generalizes our previous work about the photon's motion~\cite{JiangLin2018}.

Since the astrophysical black holes usually don't carry residual charges, the applicable scenarios of this work are mainly for the cases in which the neutral/charged particle moves in the Kerr spacetime. On the other hand, it can not be excluded the possibility that there might exist charged black holes in the Universe, and in this sense the formulations for the Kerr-Newman spacetime are more general than those for the Kerr spacetime, but the achieved formulas in this work are not applicable to the case in which both the test particle and the black hole carry charges.

The rest of paper is organized as follows. Section \ref{sec:2nd} reviews the PM dynamics equation for the test particles in Kerr-Newman spacetime. In Section \ref{sec:3rd} we derive the 2PM solutions to the trajectory and velocity of the unbound test particles. Section \ref{sec:4th} gives the corresponding formulas for two special cases. The discussions and summaries are given in Section \ref{sec:5th}.

\section{The 2PM harmonic metric and geodesic equation for the Kerr-Newman spacetime}\label{sec:2nd}

We assume that the Kerr-Newman black hole has mass $m$, electric charge $q$, and the angular momentum $\bm{J}$. The harmonic metric for the Kerr-Newman spacetime~\cite{LinJiang2014} in the 2PM approximations can be written as
{\small \begin{eqnarray}
g_{00}&=& -1 +\frac{2m}{r} -\frac{2m^2+q^2}{r^2}~,\label{eq:metric-1nd}\\
g_{0i}&=& \zeta^i~,\label{eq:metric-2nd}\\
g_{ij}&=&\Big(1 +\frac{2m}{r} + \frac{m^2}{r^2}\Big)\delta_{ij} + \frac{(m^2-q^2)x^i x^j}{r^4}~,\label{eq:metric-3nd}
\end{eqnarray}}
\hskip -0.1cm where $r\equiv |\bm{x}|$ with $\bm{x}\equiv(x^1, x^2, x^3)$ denotes the position vector of the field point, $\zeta^i$ is the $i$-th component of the gravitational vector potential $\bm{\zeta} \equiv 2 (\bm{x}\times\bm{J})/r^3$. The natural units in which the gravitational constant and the light speed in vacuum are set as $1$ are chosen, as well as the metric signature ($-+++$) with Greek indices running from 0 to 3 and Latin indices running from 1 to 3.

The motion of the test particles including the photon in general relativity is described by the geodesic equation ~\cite{Will1981,Weinberg1972,KopeikinEfroimskyKaplan2012}
{\small \begin{eqnarray}
\hskip -1.5cm \frac{d^2x^i}{dt^2} + \Gamma^i_{00} +2\Gamma^i_{0j}\frac{dx^j}{dt} +\Gamma^i_{jk}\frac{dx^j}{dt}\frac{dx^k}{dt} - \Big(\Gamma^0_{00}+ 2\Gamma^0_{0j}\frac{dx^j}{dt} +\Gamma^0_{jk}\frac{dx^j}{dt}\frac{dx^k}{dt}\Big)\frac{dx^i}{dt}=0~,\label{eq:geodesicEq-a}
\end{eqnarray}}
\hskip -0.1cm with $\Gamma_{\alpha\beta}^\mu$ being the Christoffel symbol
{\small \begin{equation}
  \Gamma_{\alpha\beta}^\mu = \frac{1}{2}g^{\rho\mu}\Big(\frac{\partial g_{\rho\beta}}{\partial x^\alpha}+\frac{\partial g_{\rho\alpha}}{\partial x^\beta} - \frac{\partial g_{\alpha\beta}}{\partial x^\rho}\Big)~.
\end{equation}}
\hskip -0.1cm Substituting Eqs.\,(\ref{eq:metric-1nd})-(\ref{eq:metric-3nd}) into Eq.\,(\ref{eq:geodesicEq-a}), we can obtain the 2PM dynamics equation of the test particles in the Kerr-Newman spacetime
{\small \begin{eqnarray}
&& \hskip -1.5cm \frac{d^2\bm{x}}{dt^2} = -\frac{m\bm{x}}{r^3} + \frac{(4m^2+q^2) \bm{x}}{r^4} + \frac{4m r - 2m^2-2q^2}{r^4}\Big(\bm{x}\!\cdot\!\frac{d\bm{x}}{dt}\Big)\frac{d\bm{x}}{dt} - \frac{(m r - q^2)\bm{x}}{r^4}\Big|\frac{d\bm{x}}{dt}\Big|^2 \nn\\
&&+ \frac{(2m^2-2q^2)\bm{x}}{r^6}\Big(\bm{x}\!\cdot\!\frac{d\bm{x}}{dt}\Big)^2 +\frac{d\bm{x}}{dt}\!\times\!(\nabla\!\times\!\bm{\zeta}) -\frac{d\bm{x}}{dt}\Big\{\frac{d\bm{x}}{dt}\!\cdot\!\Big[\Big(\!\frac{d\bm{x}}{dt}\!\cdot\!\nabla\!\Big)\bm{\zeta}\Big]\Big\}~.\label{eq:geodesicEq-2RC}
\end{eqnarray}}
\hskip -0.1cm Notice that this equation valids for both the massive particles and the massless ones.

\section{The 2PM solution for the motion of the test particles in the Kerr-Newman spacetime}\label{sec:3rd}

In this section we derive the 2PM analytical solutions to the trajectory and velocity of the test particles in the Kerr-Newman spacetime via an iterative method. We solve the geodesic equation order-by-order, similar to the procedure adopted in our previous work~\cite{JiangLin2018}. More specifically, we first give the Minkowskian solutions, and then utilize them to calculate the 1PM contributions, and then make use of the Minkowskian solutions and the 1PM contributions to further calculate the 2PM contributions.

We assume that a particle has position $\bm{x}_{\sss \rm e}$ and velocity $\bm{v}_{\sss \rm e}\equiv v_{\sss \rm e} \bm{n}$ with $\bm{n}$ being the unit vector at an initial time $t_{\sss \rm e}$. The velocity may be relativistic or non-relativistic, but it has to satisfy the small-deflection condition which is defined in Eq.\,(\ref{SDC}) below. Two small-deflection scenarios are shown in Figure \ref{Fig1}. In the first scenario (Fig.1(a)), the test particle passes by the black hole with a large impact factor. In the second scenario (Fig.1(b)), the test particle leaves away from (or moves to) some position near but not very close to the black hole, and the impact factor may be small or even zero. Here, the words ``near but not very close to the black hole" can be elaborated with a mathematical language: ``near to the black hole" means that the Newtonian potential $\phi \!\equiv\! -\frac{m}{r}$ is not enough to characterize the particle's dynamics, and the higher-order effects from the black hole's mass as well as angular momentum and charge should be taken into account in the particle's equations-of-motion; while ``not very close to the black hole" means that the effects whose magnitudes are cubic or higher orders of the Newtonian potential can be neglected. In other words, we consider all the effects from the black hole's mass and angular momentum and charge on the particle's motion, up to the quadratic order of the Newtonian potential.

\subsection{The Minkowskian solution}

Let $\bm{x}_{\sss\rm 0}$ denote the particle's position vector at any time $t\!\geq\! t_{\sss \rm e}$ in the Minkowskian spcetime. For the zeroth-order approximation, Eq.\,(\ref{eq:geodesicEq-2RC}) reduces to
{\small \begin{eqnarray}
\frac{d^2\bm{x}_{\sss\rm 0}}{dt^2} =0~,\label{eq:NewtonianEquation}
\end{eqnarray}}
\hskip -0.1cm and the Minkowskian solutions to the velocity and the trajectory of the particle can be written as
{\small \begin{eqnarray}
&&   \frac{d\bm{x}_{\sss\rm 0}}{dt} = v_e\bm{n}~,\label{NewtonianMotionV}\\
&&   \bm{x}_{\sss\rm 0} = \bm{x}_{\sss \rm e} + v_e(t -t_{\sss \rm e})\bm{n}~.\label{NewtonianMotionT}
\end{eqnarray}}

\subsection{The 1PM solution}

The 1PM solution for the motion of the test particle can be written as the sum of the Minkowskian solution and the 1PM contributions. The 1PM contributions consist of two parts: one is from the Newtonian potential, and the other is from the first-order relativistic contribution (RC).

First we consider the Newtonian term $\bm{x}_{\sss\rm N}$ contributed to the trajectory. In this case, the solution can be written as
{\small \begin{equation}
  \bm{x} = \bm{x}_{\sss\rm 0} + \bm{x}_{\sss\rm N}~,\label{eq:trajectory-N-form}
\end{equation}}

\noindent and Eq.\,(\ref{eq:geodesicEq-2RC}) reduces to
{\small \begin{equation}
\frac{d^2\bm{x}}{dt^2} = -\frac{m \bm{x}}{r^3}~.\label{eq:geodesicEq-N}
\end{equation}}

Substituting Eqs.\,(\ref{NewtonianMotionT}) and (\ref{eq:trajectory-N-form}) into Eq.\,(\ref{eq:geodesicEq-N}), and only keeping the first-order PM terms which are proportional to $m$ , we have
{\small \begin{equation}\label{eq:acceleration-N_dynamics}
  \frac{d^2\bm{x}_{\sss\rm N}}{dt^2} = -\frac{m}{\!|\bm{x}_{\sss\rm 0}|^3}\bm{x}_{\sss\rm 0}~.
\end{equation}}

\begin{figure}[h]
  \centering
  \includegraphics[width=4.5in]{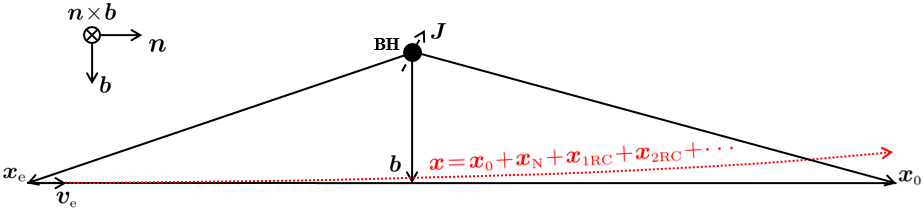}\\
  \vskip -0.2cm
  (a)\\
  \includegraphics[width=4.55in]{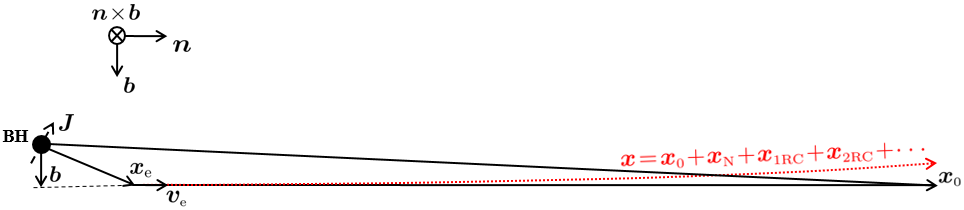}\\
    \vskip -0.2cm
  (b)\\
  \caption{Schematic diagrams for the small-deflection trajectory of the test particle in the Kerr-Newman spacetime.
  The direction of the angular momentum $\bm{J}$ of black hole is arbitrary. $\bm{x}_{\sss\rm e}$ and $\bm{v}_{\sss\rm e}$ are the initial position and velocity of the test particle. $\bm{n}$ denotes the unit vector of the initial velocity and $\bm{b}$ denotes the impact vector.
  The particle's trajectory $\bm{x} = \bm{x}_{\sss\rm 0} +\bm{x}_{\sss\rm N}+ \bm{x}_{\sss\rm 1RC}+\bm{x}_{\sss\rm 2RC}+\cdot\cdot\cdot~$ is denoted by the red dotted line and in general it is not in the incident plane spanned by $\bm{n}\!\times\!\bm{b}$. $\bm{x}_{\sss\rm 0},~\!\bm{x}_{\sss\rm N},~\!\bm{x}_{\sss\rm 1RC},~\!\bm{x}_{\sss\rm 2RC}$ represent the Minkowskian solution, the Newtonian-term contribution, the first-order and the second-order relativistic contributions to the trajectory. Notice that the 1PM contributions consist of the Newtonian-term and the first-order relativistic contribution. (a) Scenario 1: the test particle passes by the black hole with a large impact factor; (b) Scenario 2: the test particle leaves away from (or moves to) some position near but not very close to the black hole, and the impact factor may be small or even zero.  
  } \label{Fig1}
\end{figure}

In order to solve this equation conveniently, we decompose $\bm{x}_{\sss\rm N}$ into the components parallel and perpendicular to $\bm{n}$:
{\small \begin{eqnarray}
  && \bm{x}_{\sss{\rm N\parallel}} = (\bm{n}\!\cdot\!\bm{x}_{\sss\rm N})\bm{n}~, \label{eq:v-velocity-N_Parallel}\\
  && \bm{x}_{\sss\rm N\perp} = \bm{x}_{\sss\rm N} - (\bm{n}\!\cdot\!\bm{x}_{\sss\rm N})\bm{n}~.\label{eq:acceleration-N_Perp}
\end{eqnarray}}
\hskip -0.1cm Eq.\,(\ref{eq:acceleration-N_dynamics}) then yields
{\small \begin{eqnarray}
  \frac{d^2\bm{x}_{\sss\rm N\parallel}}{dt^2} &=& -\frac{m(\bm{n}\!\cdot\!\bm{x}_{\sss\rm 0})}{|\bm{x}_{\sss\rm 0}|^3}
  \bm{n}~,\label{eq:acceleration-parallel-N}\\
  \frac{d^2\bm{x}_{\sss\rm N\perp}}{dt^2} &=&
  -\frac{m}{|\bm{x}_{\sss\rm 0}|^3}\bm{b}~,\label{eq:acceleration-perp-N}
\end{eqnarray}}
\hskip -0.1cm where $\bm{b} \equiv \bm{x}_{\sss\rm 0} \!- \bm{n}(\bm{n}\!\cdot\!\bm{x}_{\sss\rm 0})=\bm{x}_{\sss\rm e} \!- \bm{n}(\bm{n}\!\cdot\!\bm{x}_{\sss\rm e})$ is the impact vector joining the center of the Kerr-Newman black hole and the point of the closest approach in the line of $\bm{x}_{\sss\rm 0}$, whose amplitude $b \equiv |\bm{b}|$ is the conventional impact parameter~\cite{Zschocke2015,JiangLin2018,Weinberg1972}. The plane spanned by the vectors $\bm{n}$ and $\bm{b}$ is defined as the incident plane. Figure \ref{Fig1} gives the schematic diagram for the trajectory of the test particle in the two different scenarios for the small-deflection motion.

Integrating Eqs.\,(\ref{eq:acceleration-parallel-N}) and (\ref{eq:acceleration-perp-N}),
we get
{\small \begin{eqnarray}\label{eq:velocity-N-perp}
 && \frac{d\bm{x}_{\sss\rm N\parallel}}{dt} = \frac{m\bm{n}}{v_e}\Big(\frac{1}{|\bm{x}_{\sss\rm 0}|} - \frac{1}{|\bm{x}_{\sss \rm e}|}\Big)~,\label{eq:velocity-N-parallel}\\
 && \frac{d\bm{x}_{\sss\rm N\perp}}{dt} = - \frac{m\bm{b}}{v_e b^2 }\Big(\frac{\bm{n}\!\cdot\!\bm{x}_{\sss\rm 0}}{|\bm{x}_{\sss\rm 0}|} -
  \frac{\bm{n}\!\cdot\!\bm{x}_{\sss \rm e}}{|\bm{x}_{\sss \rm e}|}\Big)~.\label{eq:velocity-N-perp}
\end{eqnarray}}
\hskip -0.1cm Notice that for the small-deflection approximation we have
\begin{equation}
v_e^2 \gg \frac{m}{|\bm{x}_{\sss\rm 0}|}~.\label{SDC}
\end{equation}
This condition is defined as the small-deflection condition of the test particle, and it is more stringent than the condition under which the PM method can be applied.  

Combining Eqs.\,(\ref{eq:velocity-N-parallel}) and (\ref{eq:velocity-N-perp}), we have the velocity of the particle related to the Newtonian term
{\small \begin{equation}\label{eq:velocity-particle-N}
  \frac{d\bm{x}_{\sss\rm N}}{dt} =\frac{m\bm{n}}{v_e}\Big(\frac{1}{|\bm{x}_{\sss\rm 0}|} - \frac{1}{|\bm{x}_{\sss \rm e}|}\Big) - \frac{m\bm{b}}{v_eb^2}\Big(\frac{\bm{n}\!\cdot\!\bm{x}_{\sss\rm 0}}{|\bm{x}_{\sss\rm 0}|} -
  \frac{\bm{n}\!\cdot\!\bm{x}_{\sss \rm e}}{|\bm{x}_{\sss \rm e}|}\Big)~.
\end{equation}}
\hskip -0.1cm Integrating Eq.\,(\ref{eq:velocity-particle-N}),
we can obtain the contribution from the Newtonian term to the trajectory of the particle
{\small \begin{equation}\label{eq:trajectory-particle-N}
  \bm{x}_{\sss\rm N} = -\frac{m\bm{n}}{v_e^2}\Big[\frac{\bm{n}\!\cdot\!(\bm{x}_{\sss\rm 0}\!-\!\bm{x}_{\sss\rm e})}{|\bm{x}_e|}-\ln\!{\frac{|\bm{x}_{\sss\rm 0}|\!+\!\bm{n}\!\cdot\!\bm{x}_{\sss\rm 0}}{|\bm{x}_{\sss \rm e}|\!+\!\bm{n}\!\cdot\!\bm{x}_{\sss \rm e}}}\Big] - \frac{m|\bm{x}_{\sss\rm 0}|\bm{b}}{v_e^2\,b^2}\Big(1 \!-\! \frac{\bm{x}_{\sss \rm e}\!\cdot\!\bm{x}_{\sss\rm 0}}{|\bm{x}_{e}||\bm{x}_{\sss\rm 0}|}\Big).
\end{equation}}

Next we include the the first-order relativistic contribution which is comparable to the Newtonian term only when $v_e$ is highly relativistic, and write the full 1PM solution as follow
{\small \begin{equation}\label{eq:trajectory-1RC-general}
  \bm{x} = \bm{x}_{\sss\rm 0} + \bm{x}_{\sss\rm N}+ \bm{x}_{\sss\rm 1RC}~,
\end{equation}}
\noindent with $\bm{x}_{\sss\rm 1RC}$ being the first-order relativistic contribution.

Substituting Eq.\,(\ref{eq:trajectory-1RC-general}) into Eq.\,(\ref{eq:geodesicEq-2RC}), and making use of Eqs.\,(\ref{NewtonianMotionT}),\,(\ref{eq:acceleration-N_dynamics}),\,(\ref{eq:velocity-particle-N})-(\ref{eq:trajectory-particle-N}), we can obtain
{\small \begin{equation}\label{eq:acceleration-1RC_dynamics}
  \frac{d^2\bm{x}_{\sss\rm 1RC}}{dt^2} = -v_e^2\frac{m\bm{x}_{\sss\rm 0}}{\!|\bm{x}_{\sss\rm 0}|^3}+v_e^2\frac{4m}{|\bm{x}_{\sss\rm 0}|^3}\bm{n}(\bm{n}\!\cdot\!\bm{x}_{\sss\rm 0})~,
\end{equation}}
\noindent here the terms higher than the 1PM order have been dropped.

Similarly, we can decompose $\bm{x}_{\sss\rm 1RC}$ into the components parallel and perpendicular to $\bm{n}$:
{\small
\begin{eqnarray}
 && \bm{x}_{\sss\rm 1RC\parallel} =(\bm{n}\!\cdot\!\bm{x}_{\sss\rm 1RC}) \bm{n}~, \label{1PNparallel}\\
 && \bm{x}_{\sss\rm 1RC\perp} = \bm{x}_{\sss\rm 1RC} - (\bm{n}\!\cdot\!\bm{x}_{\sss\rm 1RC}) \bm{n}~.\label{1PNperp}
\end{eqnarray}
}
\hskip -0.1cm From Eqs.\,(\ref{eq:acceleration-1RC_dynamics}),\,(\ref{1PNparallel}) and (\ref{1PNperp}) we have
{\small \begin{eqnarray}
\frac{d^2\bm{x}_{\sss\rm 1RC\parallel}}{dt^2} &=& \frac{3mv_e^2}{|\bm{x}_{\sss\rm 0}|^3}(\bm{n}\!\cdot\!\bm{x}_{\sss\rm 0})\bm{n}~,\label{eq:acceleration-parallel-1RC}\\
  \frac{d^2\bm{x}_{\sss\rm 1RC\perp}}{dt^2} &=& -\frac{mv_e^2}{|\bm{x}_{\sss\rm 0}|^3}\bm{b}~.\label{eq:acceleration-perp-1RC}
\end{eqnarray}}
\hskip -0.1cm Integrating Eqs.\,(\ref{eq:acceleration-parallel-1RC}) and (\ref{eq:acceleration-perp-1RC}),
we can obtain the first-order relativistic contribution to the velocity
{\small \begin{eqnarray}
&&\frac{d\bm{x}_{\sss\rm 1RC}}{dt} =
3v_e\bm{n}\Big(\frac{m}{|\bm{x}_{\sss \rm e}|} - \frac{m}{|\bm{x}_{\sss\rm 0}|} \Big) - v_e\frac{m\bm{b}}{b^2}\Big(\frac{\bm{n}\!\cdot\!\bm{x}_{\sss\rm 0}}{|\bm{x}_{\sss\rm 0}|} -
  \frac{\bm{n}\!\cdot\!\bm{x}_{\sss \rm e}}{|\bm{x}_{\sss \rm e}|}\Big)~.\label{eq:velocity-particle-1RC}
\end{eqnarray}}
\hskip -0.1cm The first-order relativistic contribution $\bm{x}_{\sss\rm 1RC}$ to the trajectory can be achieved via integrating Eq.\,(\ref{eq:velocity-particle-1RC})
as follow
{\small \begin{eqnarray}
&&\bm{x}_{\sss\rm 1RC} = 3m\bm{n}\Big[\frac{\bm{n}\!\cdot\!(\bm{x}_{\sss\rm 0}\!-\!\bm{x}_{\sss\rm e})}{|\bm{x}_{\sss\rm e}|}-\ln\!{\frac{|\bm{x}_{\sss\rm 0}|\!+\!\bm{n}\!\cdot\!\bm{x}_{\sss\rm 0}}{|\bm{x}_{\sss \rm e}|\!+\!\bm{n}\!\cdot\!\bm{x}_{\sss \rm e}}}\Big] - \frac{m|\bm{x}_{\sss\rm 0}|\bm{b}}{b^2}\Big(1 - \frac{\bm{x}_{\sss \rm e}\!\cdot\!\bm{x}_{\sss\rm 0}}{|\bm{x}_{e}||\bm{x}_{\sss\rm 0}|}\Big)~.\label{eq:trajectory-particle-1RC}
\end{eqnarray}}


\subsection{The 2PM solution}
Following the same procedure, we can further deduce the 2PM trajectory and velocity for the test particle using the iterative method.
To the 2PM-order accuracy, the trajectory and velocity solutions of Eq.\,(\ref{eq:geodesicEq-2RC}) can be written as
{\small \begin{eqnarray}
  \bm{x} = \bm{x}_{\sss\rm 0} +\bm{x}_{\sss\rm N}+ \bm{x}_{\sss\rm 1RC}+ \bm{x}_{\sss\rm 2RC}~,\label{eq:trajectory-2RC-general}
\end{eqnarray}
 \begin{eqnarray}
  \frac{d\bm{x}}{dt} = \frac{d\bm{x}_{\sss\rm 0}}{dt} +\frac{d\bm{x}_{\sss\rm N}}{dt}+ \frac{d\bm{x}_{\sss\rm 1RC}}{dt}+ \frac{d\bm{x}_{\sss\rm 2RC}}{dt}~,\label{eq:velocity-2RC-general}
\end{eqnarray}}
\noindent \hskip -0.15cm with $\bm{x}_{\sss\rm 2RC}$ and $\frac{d\bm{x}_{\sss\rm 2RC}}{dt}$ being the second-order relativistic contributions to the trajectory and velocity respectively.

Similarly, we can decompose $\bm{x}_{\sss\rm 2RC}$ into the components parallel and perpendicular to $\bm{n}$:
{\small \begin{equation}
  \bm{x}_{\sss\rm 2RC} = \bm{x}_{\sss\rm 2RC\parallel} + \bm{x}_{\sss\rm 2RC\perp} \label{2RCTrajectory}
\end{equation}}
\noindent with{\small
\begin{eqnarray}
 && \bm{x}_{\sss\rm 2RC\parallel} = (\bm{n}\!\cdot\!\bm{x}_{\sss\rm 2RC})\bm{n}~, \label{2PNparallel}\\
 && \bm{x}_{\sss\rm 2RC\perp} = \bm{x}_{\sss\rm 2RC} - (\bm{n}\!\cdot\!\bm{x}_{\sss\rm 2RC})\bm{n}~.\label{2PNperp}
\end{eqnarray}}
\noindent \hskip -0.1cm Substituting Eqs.\,(\ref{eq:trajectory-2RC-general})-(\ref{2PNperp}) into Eq.\,(\ref{eq:geodesicEq-2RC}), making use of Eqs.\,(\ref{NewtonianMotionV}),\,(\ref{NewtonianMotionT}),\,(\ref{eq:velocity-particle-N}),\,(\ref{eq:trajectory-particle-N}),\,(\ref{eq:acceleration-1RC_dynamics}),\, (\ref{eq:velocity-particle-1RC}) and (\ref{eq:trajectory-particle-1RC}), we can obtain
{\small \begin{eqnarray}
 && \hskip -2cm \frac{d^2\bm{x}_{\sss\rm 2RC\parallel}}{dt^2} \!=\!\frac{m^2\bm{n}}{|\bm{x}_{\sss\rm 0}|^3}\Big\{\!\Big(\!9v_e^2\!-\!6\!+\!\frac{1}{v_e^2}\!\Big)\!\Big(\!2\!-\!\frac{3b^2}{|\bm{x}_{\sss\rm 0}|^2}\!\Big)\!\Big[\!\ln\!{\frac{|\bm{x}_{\sss\rm 0}|\!+\!\bm{n}\!\cdot\!\bm{x}_{\sss\rm 0}}{|\bm{x}_{\sss \rm e}|\!+\!\bm{n}\!\cdot\!\bm{x}_{\sss \rm e}}}\!  -\! v_e \!\frac{(t-t_e)}{|\bm{x}_{\sss\rm e}|}\Big]\!-\!\Big(\!5v_e^2\!+\!2\!-\!\frac{3}{v_e^2}\!\Big)\!\frac{\bm{n}\!\cdot\!\bm{x}_{\sss\rm e}}{|\bm{x}_{\sss\rm e}|}
\nn\\
&&  \hskip 0.85cm \!+\!\Big[\!\Big(3v_e^2\!-\!1\Big)\!\Big(\frac{6|\bm{x}_{\sss\rm 0}|}{|\bm{x}_{\sss\rm e}|}\!-\!\frac{q^2}{m^2}\!\Big)\!-\!\Big(\!13v_e^2\!-\!12\!+\!\frac{3}{v_e^2}\!\Big)\!-\!\frac{2v_e^2(m^2\!-\!q^2)b^2}{m^2|\bm{x}_{\sss\rm 0}|^2}\Big]\!\frac{\bm{n}\!\cdot\!\bm{x}_{\sss\rm 0}}{|\bm{x}_{\sss\rm 0}|}\! \nn\\
 && \hskip 0.85cm \!-\!\Big(\!9v_e^2\!+\!6\!-\!\frac{3}{v_e^2}\!\Big)\!\frac{b^2\bm{n}\!\cdot\!(\bm{x}_{\sss\rm 0}\!-\!\bm{x}_{\sss\rm e})}{|\bm{x}_{\sss\rm 0}|^2|\bm{x}_{\sss\rm e}|} \!+ \!\frac{v_e^3|\bm{x}_{\sss\rm 0}|^3\bm{n}\!\cdot\![(\bm{n}\!\cdot\!\nabla)\bm{\zeta}]}{m^2}\Big\},~~\label{eq:acceleration-parallel-2RC}
 \end{eqnarray}
 \begin{eqnarray}
&& \hskip -2cm \frac{d^2\bm{x}_{\sss\rm 2RC\perp}}{dt^2} \!=\! \frac{m^2\bm{b}}{|\bm{x}_{\sss\rm 0}|^3}\Big\{\!\Big(\!-\!18\!+\!\frac{6}{v_e^2}\!\Big)\!\frac{\bm{n}\!\cdot\!\bm{x}_{\sss\rm 0}}{|\bm{x}_{\sss\rm 0}|^2}\!\ln\!{\frac{|\bm{x}_{\sss\rm 0}|\!+\!\bm{n}\!\cdot\!\bm{x}_{\sss\rm 0}}{|\bm{x}_{\sss \rm e}|\!+\!\bm{n}\!\cdot\!\bm{x}_{\sss \rm e}}}\!+\!\Big(\!9v_e^2\!-\!\frac{3}{v_e^2}\!\Big)\!\frac{1}{|\bm{x}_{\sss\rm 0}|}\!-\!\Big(\!10v_e^2\!-\!16\!+\!\frac{6}{v_e^2}\Big)\!\frac{1}{|\bm{x}_{\sss \rm e}|} \nn\\
&&\hskip 0.9cm \!-\!\Big(\!3v_e^2\!+\!2\!-\!\frac{1}{v_e^2}\!\Big)\!\frac{|\bm{x}_{\sss\rm 0}|}{b^2}\!+\!\Big[\Big(\!3v_e^2\!-\!12\!+\!\frac{9}{v_e^2}\!\Big)\!\frac{1}{|\bm{x}_{\sss\rm 0}|^2}\!+\!\Big(\!3v_e^2\!+\!2\!-\!\frac{1}{v_e^2}\!\Big)\!\frac{1}{b^2}\Big]\!\frac{\bm{x}_{\sss\rm e}\!\cdot\!\bm{x}_{\sss\rm 0}}{|\bm{x}_{\sss\rm e}|}\nn\\
&&\hskip 0.9cm \!-\frac{2b^2v_e^2}{|\bm{x}_{\sss\rm 0}|^3}-\frac{(v_e^2-1)q^2}{m^2|\bm{x}_{\sss\rm 0}|}+\frac{2q^2b^2v_e^2}{m^2|\bm{x}_{\sss\rm 0}|^3}\Big\}+\bm{v_e}\!\times\!(\nabla\!\times\!\bm{\zeta})\!~.\label{eq:acceleration-perp-2RC}
\end{eqnarray}}
\noindent \hskip -0.1cm where the terms higher than the 2PM order have been dropped. Here and from now on the vector potential $\bm{\zeta}$ is evaluated by $2(\bm{x}_{\sss\rm 0}\!\times\!\bm{J})/|\bm{x}_{\sss\rm 0}|^3$. In the following integration we will use the identity of  {\small $\bm{x}_{\sss\rm e}\!\cdot\!\bm{x}_{\sss\rm 0}\!=\!b^2\!+\!(\bm{n}\!\cdot\!\bm{x}_{\sss\rm e})(\bm{n}\!\cdot\!\bm{x}_{\sss\rm 0})$~}.

Integrating Eqs.\,(\ref{eq:acceleration-parallel-2RC}) and (\ref{eq:acceleration-perp-2RC}),
we can obtain the second-order relativistic contribution to the velocity
{\small \begin{eqnarray}
\frac{d\bm{x}_{\sss\rm 2RC}}{dt} = \frac{d\bm{x}_{\sss\rm 2RC\parallel}}{dt} + \frac{d\bm{x}_{\sss\rm 2RC\perp}}{dt}~,\label{eq:velocity-2PM}
\end{eqnarray}}
\hskip -0.1cm with
{\small \begin{eqnarray}
 &&  \hskip -2cm \frac{d\bm{x}_{\sss\rm 2RC\parallel}}{dt}
=v_em^2\bm{n}\Big\{\!\Big(\!\!-\!9\!+\!\frac{6}{v_e^2}\!-\!\frac{1}{v_e^4}\!\Big)
\!\frac{\bm{n}\!\cdot\!\bm{x}_{\sss\rm 0}}{|\bm{x}_{\sss\rm 0}|^3}\!\ln\!{\frac{|\bm{x}_{\sss\rm 0}|\!+\!\bm{n}\!\cdot\!\bm{x}_{\sss\rm 0}}{|\bm{x}_{\sss \rm e}|\!+\!\bm{n}\!\cdot\!\bm{x}_{\sss \rm e}}}-\Big(\!\frac{6}{v_e^2}\!-\!\frac{2}{v_e^4}\!\Big)\!\frac{1}{|\bm{x}_{\sss\rm e}|}\!\Big(\!\frac{1}{|\bm{x}_{\sss\rm 0}|}\!-\!\frac{1}{|\bm{x}_{\sss\rm e}|}\Big)
\nn\\
 && \hskip 1.1cm \!+\Big[\Big(\!2\!-\!\frac{3}{v_e^2}\!+\!\frac{1}{v_e^4}\!\Big)\!
+\!\Big(\!\frac{3}{2}\!-\!\frac{1}{2v_e^2}\!\Big)\!\frac{q^2}{m^2}\Big]\!\Big(\!\frac{1}{|\bm{x}_{\sss\rm 0}|^2}\!-\!\frac{1}{|\bm{x}_{\sss\rm e}|^2}\!\Big)\!\nn\\
 && \hskip 1.1cm \!+\frac{(m^2\!-\!q^2)b^2}{2m^2}\!\Big(\frac{1}{|\bm{x}_{\sss\rm 0}|^4}\!-\!\frac{1}{|\bm{x}_{\sss\rm e}|^4}\Big)\!+\!\Big(\!1\!+\!\frac{2}{v_e^2}\!+\!\frac{1}{v_e^4}\Big)
\!\frac{\bm{n}\!\cdot\!\bm{x}_{\sss\rm e}}{b^2|\bm{x}_{\sss\rm e}|}\!\Big(\frac{\bm{n}\!\cdot\!\bm{x}_{\sss\rm 0}}{|\bm{x}_{\sss\rm 0}|}\!-\!\frac{\bm{n}\!\cdot\!\bm{x}_{\sss\rm e}}{|\bm{x}_{\sss\rm e}|}\Big)\!\nn\\
 && \hskip 1.1cm \!-\!\Big(\!6\!-\!\frac{8}{v_e^2}
\!+\!\frac{2}{v_e^4}\!\Big)
\!\Big(\!\frac{\bm{x}_{\sss\rm e}\!\cdot\!\bm{x}_{\sss\rm 0}}{|\bm{x}_{\sss\rm e}||\bm{x}_{\sss\rm 0}|^3}\!-\!\frac{1}{|\bm{x}_{\sss\rm e}|^2}\!\Big)\!-\!\frac{2v_e^2\bm{n}\!\cdot\!(\bm{x}_{\sss\rm 0}\!\times\!\bm{J})}{m^2|\bm{x}_{\sss\rm 0}|^3}
\Big\}.\label{eq:velocity-2PM-para}
\end{eqnarray}}
{\small \begin{eqnarray}
 && \hskip -2cm \frac{d\bm{x}_{\sss\rm 2RC\perp}}{dt}
= \bm{n}\!\times\!\bm{b}\Big\{\!2\bm{n}\!\cdot\!\bm{J}\Big(\frac{1}{|\bm{x}_{\sss\rm 0}|^3}\!-\!\frac{1}{|\bm{x}_{\sss\rm e}|^3}\Big)\!
-\!\frac{2\bm{b}\!\cdot\!\bm{J}}{b^2}\!\Big(\frac{\bm{n}\!\cdot\!\bm{x}_{\sss\rm 0}}{b^2|\bm{x}_{\sss\rm 0}|}\!-\!\frac{\bm{n}\!\cdot\!\bm{x}_{\sss\rm e}}{b^2|\bm{x}_{\sss\rm e}|}\!+\!\frac{\bm{n}\!\cdot\!\bm{x}_{\sss\rm 0}}{|\bm{x}_{\sss\rm 0}|^3}\!-\!\frac{\bm{n}\!\cdot\!\bm{x}_{\sss\rm e}}{|\bm{x}_{\sss\rm e}|^3}\Big)\!\Big\} \nn\\
 && \hskip -0.35cm \!+v_e\bm{b}\Big\{\!\Big(\!\frac{6}{v_e^2}\!-\!\frac{2}{v_e^4}\!\Big)\!\frac{m^2}{|\bm{x}_{\sss\rm 0}|^3}\!\ln\!{\frac{|\bm{x}_{\sss\rm 0}|\!+\!\bm{n}\!\cdot\!\bm{x}_{\sss\rm 0}}{|\bm{x}_{\sss \rm e}|\!+\!\bm{n}\!\cdot\!\bm{x}_{\sss \rm e}}}\!-\!\Big(\!5\!-\!\frac{10}{v_e^2}\!+\!\frac{1}{v_e^4}\Big)\!\frac{m^2}{b^2|\bm{x}_{\sss\rm e}|}\!\Big(\!\frac{\bm{n}\!\cdot\!\bm{x}_{\sss\rm 0}}{|\bm{x}_{\sss\rm 0}|}\!-\!\frac{\bm{n}\!\cdot\!\bm{x}_{\sss\rm e}}{|\bm{x}_{\sss\rm e}|}\!\Big)\nn\\
 && \hskip 0.85cm \!-\!\Big[\!\Big(\!\frac{3}{4}\!-\!\frac{5}{v_e^2}\!+\!\frac{1}{2v_e^4}\!\Big)\!\frac{m^2}{b^3}\!+\!\Big(\frac{1}{4}\!+\!\frac{1}{2v_e^2}\!\Big)
\!\frac{q^2}{b^3}\Big]\!\Big(\!\arccos\frac{\bm{n}\!\cdot\!\bm{x}_{\sss\rm 0}}{|\bm{x}_{\sss\rm 0}|}\!-\!\arccos\frac{\bm{n}\!\cdot\!\bm{x}_{\sss\rm e}}{|\bm{x}_{\sss\rm e}|}\!\Big)\nn\\
 && \hskip 0.85cm \!+\!\Big(\!1\!-\!\frac{4}{v_e^2}\!+\!\frac{3}{v_e^4}\!\Big)\!\frac{m^2\bm{n}\!\cdot\!(\bm{x}_{\sss\rm 0}\!-\!\bm{x}_{\sss\rm e})}{|\bm{x}_{\sss\rm e}||\bm{x}_{\sss\rm 0}|^3}\!-\!\Big(\!3\!+\!\frac{2}{v_e^2}\!-\!\frac{1}{v_e^4}\!\Big)\!\frac{m^2(\bm{n}\!\cdot\!\bm{x}_{\sss\rm e})}{b^2|\bm{x}_{\sss\rm e}|}\!\Big(\!\frac{1}{|\bm{x}_{\sss\rm 0}|}\!-\!\frac{1}{|\bm{x}_{\sss\rm e}|}\!\Big)\nn\\
 && \hskip 0.85cm \!+\!\Big[\!\Big(\!\frac{15}{4}\!-\!\frac{3}{v_e^2}\!-\!\frac{1}{2v_e^4}\Big)\!\frac{m^2}{b^2}\!
 +\!\Big(\frac{1}{4}\!+\!\frac{1}{2v_e^2}\Big)\!\frac{q^2}
{b^2}\Big]\!\Big(\!\frac{\bm{n}\!\cdot\!\bm{x}_{\sss\rm 0}}{|\bm{x}_{\sss\rm 0}|^2}\!-\!\frac{\bm{n}\!\cdot\!\bm{x}_{\sss\rm e}}{|\bm{x}_{\sss\rm e}|^2}\!\Big)\!\nn\\
 && \hskip 0.85cm \!-\!\Big(\!\frac{m^2}{2}\!-\!\frac{q^2}{2}\Big)\!\Big(\!\frac{\bm{n}\!\cdot\!\bm{x}_{\sss\rm 0}}{|\bm{x}_{\sss\rm 0}|^4}\!-\frac{\bm{n}\!\cdot\!\bm{x}_{\sss\rm e}}{|\bm{x}_{\sss\rm e}|^4}\Big)\!-\!\frac{2(\bm{n}\!\times\!\bm{b})\!\cdot\!\bm{J}}{b^4v_e}\!\Big(\!\frac{\bm{n}\!\cdot\!\bm{x}_{\sss\rm 0}}{|\bm{x}_{\sss\rm 0}|}\!-\!\frac{\bm{n}\!\cdot\!\bm{x}_{\sss\rm e}}{|\bm{x}_{\sss\rm e}|}\Big)\!\Big\}.\label{eq:velocity-2PM-perp}
\end{eqnarray}}
\noindent \hskip -0.1cm The second-order relativistic contribution $\bm{x}_{\sss\rm 2RC}$ to the trajectory can be achieved via integrating Eqs\,(\ref{eq:velocity-2PM-para}) and (\ref{eq:velocity-2PM-perp})
{\small \begin{eqnarray}
 &&\hskip -2cm \bm{x}_{\sss\rm 2RC\parallel}
\!=\bm{n}\frac{m^2}{b}\Big\{\!\Big[\!\Big(\!\frac{27}{4}\!-\!\frac{3}{v_e^2}\!\Big)\!-\!
\Big(\!\frac{5}{4}\!-\!\frac{1}{2v_e^2}\!\Big)\!\frac{q^2}{m^2}
\Big]\!\Big(\arccos\frac{\bm{n}\!\cdot\!\bm{x}_{\sss\rm 0}}{|\bm{x}_{\sss\rm 0}|}\!-\!\arccos\frac{\bm{n}\!\cdot\!\bm{x}_{\sss\rm e}}{|\bm{x}_{\sss\rm e}|}\Big)\nn\\
 && \hskip -0.4cm  +\!\Big[\Big(\!9\!-\!\frac{6}{v_e^2}\!+\!\frac{1}{v_e^4}\!\Big)\!\frac{b}{|\bm{x}_{\sss\rm 0}|}\!-\!\Big(\!\frac{6}{v_e^2}\!-\!\frac{2}{v_e^4}\!\Big)\!\frac{b}{|\bm{x}_{\sss\rm e}|}\Big]\!\ln\!{\frac{|\bm{x}_{\sss\rm 0}|\!+\!\bm{n}\!\cdot\!\bm{x}_{\sss\rm 0}}{|\bm{x}_{\sss \rm e}|\!+\!\bm{n}\!\cdot\!\bm{x}_{\sss \rm e}}}\!+\!\frac{b(m^2\!-\!q^2)}{4m^2}\!\Big(\!\frac{\bm{n}\!\cdot\!\bm{x}_{\sss\rm 0}}{|\bm{x}_{\sss\rm 0}|^2}\!-\!\frac{\bm{n}\!\cdot\!\bm{x}_{\sss\rm e}}{|\bm{x}_{\sss\rm e}|^2}\!\Big)\!\nn\\
 && \hskip -0.4cm +\!\Big[\!\Big(\!\frac{11}{2}\!-\!\frac{8}{v_e^2}\!+\!\frac{2}{v_e^4}\!\Big)\!\frac{b^2}{|\bm{x}_{\sss\rm e}|^2}\!+\!\frac{q^2b^2}{2m^2|\bm{x}_{\sss\rm e}|^2}\!-\!\Big(\!2\!-\!\frac{9}{v_e^2}\!+\!\frac{3}{v_e^4}\!\Big)\!
-\!\Big(\!\frac{3}{2}\!
-\!\frac{1}{2v_e^2}\!\Big)\!\frac{q^2}{m^2}\!\Big]\!\frac{b\,\bm{n}\!\cdot\!(\bm{x}_{\sss\rm 0}\!-\!\bm{x}_{\sss\rm e})}{|\bm{x}_{\sss\rm e}|^2}\nn\\
 && \hskip -0.4cm+\!\Big[\!\Big(\!1\!+\!\frac{2}{v_e^2}\!+\!\frac{1}{v_e^4}\!\Big)\!\Big(\!\frac{|\bm{x}_{\sss\rm 0}|}{b^2}-\frac{\bm{x}_{\sss\rm e}\!\cdot\!\bm{x}_{\sss\rm 0}}{|\bm{x}_{\sss\rm e}|b^2}\Big)\!+\!\Big(\!6\!-\!\frac{8}{v_e^2}\!+\!\frac{2}{v_e^4}\!\Big)\!\Big(\!\frac{1}{|\bm{x}_{\sss\rm 0}|}\!-\!\frac{2}{|\bm{x}_{\sss\rm e}|}\!+\!\frac{\bm{x}_{\sss\rm e}\!\cdot\!\bm{x}_{\sss\rm 0}}{|\bm{x}_{\sss\rm e}|^3}\Big)\!\Big]\!\frac{b(\bm{n}\!\cdot\!\bm{x}_{\sss\rm e})}{|\bm{x}_{\sss\rm e}|}\nn\\
 && \hskip -0.4cm -\!\Big[\!\Big(\!6\!-\!\frac{8}{v_e^2}\!+\!\frac{2}{v_e^4}\!\Big)\!\frac{b}{|\bm{x}_{\sss\rm e}|}\!+\!\frac{2v_e^2(\bm{n}\!\times\!\bm{b})\!\cdot\!\bm{J}}{b\,m^2}\Big]\!\Big(\!\frac{\bm{n}\!\cdot\!\bm{x}_{\sss\rm 0}}{|\bm{x}_{\sss\rm 0}|}\!-\frac{\bm{n}\!\cdot\!\bm{x}_{\sss\rm e}}{|\bm{x}_{\sss\rm e}|}\!\Big)\!\Big\}.\label{2RCTrajectory-para}
\end{eqnarray}}
{\small \begin{eqnarray}
 &&\hskip -2cm  \bm{x}_{\sss\rm 2RC\perp}
\!=\bm{b}\frac{m^2}{b^2}\Big\{\Big[\!\Big(\frac{6}{v_e^2}\!-\!\frac{2}{v_e^4}\!\Big)\!\frac{\bm{n}\!\cdot\!\bm{x}_{\sss\rm 0}}{|\bm{x}_{\sss\rm 0}|}\!-\!\Big(\!3\!+\!\frac{2}{v_e^2}\!-\!\frac{1}{v_e^4}\!\Big)\!\frac{\bm{n}\!\cdot\!\bm{x}_{\sss\rm e}}{|\bm{x}_{\sss\rm e}|}\Big]\!\ln\!{\frac{|\bm{x}_{\sss\rm 0}|\!+\!\bm{n}\!\cdot\!\bm{x}_{\sss\rm 0}}{|\bm{x}_{\sss \rm e}|\!+\!\bm{n}\!\cdot\!\bm{x}_{\sss \rm e}}}\!+\!\Big(\!3\!-\!\frac{4}{v_e^2}\!+\!\frac{1}{v_e^4}\!\Big)\!\ln\!\frac{|\bm{x}_{\sss\rm 0}|}{\!|\bm{x}_{\sss\rm e}|\!}\!\nn\\
&& \hskip 0.35cm -\!\Big[\!\Big(\frac{3}{4}\!-\!\frac{5}{v_e^2}\!+\!\frac{1}{2v_e^4}\!\Big)\!+\!\Big(\!\frac{1}{4}\!+\!\frac{1}{2v_e^2}\!\Big)
\!\frac{q^2}{m^2}\Big]\!\frac{\bm{n}\!
\cdot\!\bm{x}_{\sss\rm 0}}{b}\!\Big(\!\arccos\frac{\bm{n}\!\cdot\!\bm{x}_{\sss\rm 0}}{|\bm{x}_{\sss\rm 0}|}\!-\!\arccos\frac{\bm{n}\!\cdot\!\bm{x}_{\sss\rm e}}{|\bm{x}_{\sss\rm e}|}\!\Big)\nn\\
&& \hskip 0.35cm
+\!\Big(\frac{1}{4}\!-\!\frac{q^2}{4m^2}\!\Big)\!\Big(\!\frac{b^2}{|\bm{x}_{\sss\rm 0}|^2}\!-\!\frac{b^2}{|\bm{x}_{\sss\rm e}|^2}\!\Big)\!
+\!\Big[\!\Big(\!8\!-\!\frac{8}{v_e^2}\!\Big)\!\frac{|\bm{x}_{\sss\rm 0}|}{|\bm{x}_{\sss\rm e}|}\!-\!\Big(\!1\!-\!\frac{4}{v_e^2}\!+\!\frac{3}{v_e^4}\Big)\!\Big]\!\frac{\bm{x}_{\sss\rm e}\!\cdot\!\bm{x}_{\sss\rm 0}}{|\bm{x}_{\sss\rm e}||\bm{x}_{\sss\rm 0}|}\nn\\
&& \hskip 0.35cm
+\!\Big[\!\Big(\!\frac{15}{4}\!-\!\frac{3}{v_e^2}\!-\!\frac{1}{2v_e^4}\!\Big)\!
+\!\Big(\frac{1}{4}\!+\!\frac{1}{2v_e^2}\!\Big)\!\frac{q^2}{m^2}\!-\!\Big(\!\frac{1}{2}
\!-\!\frac{q^2}{2m^2}\!\Big)\!\frac{b^2}{|\bm{x}_{\sss\rm e}|^2}\Big]\!\Big(\!1\!-\!\frac{\bm{x}_{\sss\rm e}\!\cdot\!\bm{x}_{\sss\rm 0}}{|\bm{x}_{\sss\rm e}|^2}\!\Big)\nn\\
&& \hskip  0.35cm
-\!\Big(\!2\!+\!\frac{6}{v_e^2}-\frac{4}{v_e^4}\!\Big)\!-\!\Big(\!5\!-\!\frac{10}{v_e^2}\!+\!\frac{1}{v_e^4}\!\Big)\!\frac{|\bm{x}_{\sss\rm 0}|}{|\bm{x}_{\sss\rm e}|}\!-\!\frac{2|\bm{x}_{\sss\rm 0}|(\bm{n}\!\times\!\bm{b})\!\cdot\!\bm{J}}{v_e\,b^2\,m^2}\!\Big(\!1\!-\frac{\!\bm{x}_{\sss\rm e}\!\cdot\!\bm{x}_{\sss\rm 0}}{|\bm{x}_{\sss\rm e}||\bm{x}_{\sss\rm 0}|}\Big)\!\Big\}\nn\\
&&  \hskip -1cm +\bm{n}\!\times\!\bm{b}\Big\{\frac{2\bm{n}\!\cdot\!\bm{J}}{b^2v_e}\!\Big[\frac{\bm{n}\!\cdot\!\bm{x}_{\sss\rm 0}}{|\bm{x}_{\sss\rm 0}|}\!-\!\frac{\bm{n}\!\cdot\!\bm{x}_{\sss\rm e}}{|\bm{x}_{\sss\rm e}|}\!-\!\frac{b^2\bm{n}\!\cdot\!(\bm{x}_{\sss\rm 0}\!-\!\bm{x}_{\sss\rm e})}{|\bm{x}_{\sss\rm e}|^3}\Big]\! \nn\\
&&  \hskip 0.5cm +\!\frac{2\bm{b}\!\cdot\!\bm{J}}{b^2v_e}\!\Big[\frac{1}{|\bm{x}_{\sss\rm 0}|}\!-\!\frac{2}{|\bm{x}_{\sss\rm e}|}\!-\!\frac{|\bm{x}_{\sss\rm 0}|}{b^2}\!+\!\frac{\bm{x}_{\sss\rm e}\!\cdot\!\bm{x}_{\sss\rm 0}}{b^2|\bm{x}_{\sss\rm e}|}\!\Big(\!1\!+\!\frac{b^2}{|\bm{x}_{\sss\rm e}|^2}\!\Big)\!\Big]\Big\}~
.\label{2RCTrajectory-perp}
\end{eqnarray}}

For readers' convenience, some integrals used in the derivations are given in~\ref{Integrals}.

It is worth pointing out that the above formulas valid for the cases in which the particle leaves away from (or moves to) some position near to the black hole with $v_e^2 \!\gg\! m/|\bm{x}_{\sss\rm 0}|$ and any impact factor. In the limit of $b\rightarrow 0$, we can expanded the above equations into the powers of $b$, and arrive at the following solutions
{\small \begin{eqnarray}
 && \hskip -2cm \frac{d^2\bm{x}}{dt^2}
\!=\!\frac{m\bm{n}}{|\bm{x}_{\sss\rm 0}|^2}\Big\{\Big(3v_e^2\!-\!1\Big)\!\Big(\!1\!-\!\frac{q^2}{m|\bm{x}_{\sss\rm 0}|}\!\Big)\!+\!\frac{m}{|\bm{x}_{\sss\rm 0}|}\!\Big[\!\Big(\!18v_e^2\!-\!12\!+\!\frac{2}{v_e^2}\!\Big)\!\ln\!\frac{|\bm{x}_{\sss\rm 0}|}{|\bm{x}_{\sss\rm e}|}\!-\!\Big(\!2\!-\!\frac{2}{v_e^2}\!\Big)\!+\!\Big(\!6\!-\!\frac{2}{v_e^2}\!\Big)\!\frac{|\bm{x}_{\sss\rm 0}|}{|\bm{x}_{\sss\rm e}|}\Big]\Big\}\nn\\
&& \hskip -1.6cm +\bm{b}\frac{m^2}{|\bm{x}_{\sss\rm 0}|^4}\Big\{\!-\!(v_e^2\!+\!1)\frac{|\bm{x}_{\sss\rm 0}|}{m}\!+\!\Big(\!9v_e^2\!-\!\frac{3}{v_e^2}\!\Big)\!-\!\Big(\!18\!-\!\frac{6}{v_e^2}\!\Big)\!\ln\!{\frac{|\bm{x}_{\sss\rm 0}|}{|\bm{x}_{\sss \rm e}|}}\!-\!\Big(\!\frac{3v_e^2}{2}\!+\!1\!-\!\frac{1}{2v_e^2}\!\Big)\!\Big(\!1\!-\!\frac{|\bm{x}_{\sss\rm 0}|}{|\bm{x}_{\sss\rm e}|}\Big)^{\!2}\!\nn\\
&&\hskip -1.3cm -\!\Big(\!10v_e^2\!-\!16\!+\!\frac{6}{v_e^2}\!\Big)\!\frac{|\bm{x}_{\sss\rm 0}|}{|\bm{x}_{\sss \rm e}|}\!+\!\Big(\!3v_e^2\!-\!12\!+\!\frac{9}{v_e^2}\!\Big)\!\frac{\bm{x}_{\sss\rm e}\!\cdot\!\bm{x}_{\sss\rm 0}}{|\bm{x}_{\sss\rm e}||\bm{x}_{\sss\rm 0}|}\!-\!\frac{(v_e^2\!-\!1)q^2}{m^2} \!-\!\frac{2|\bm{x}_{\sss\rm 0}|(\bm{n}\!\times\!\bm{b})\!\cdot\!\bm{J}}{v_em^2b^2}\!\Big\}\nn\\
&& \hskip -1.6cm  + v_e(\bm{n}\!\times\!\bm{b})\Big\{\frac{2\bm{b}\!\cdot\!\bm{J}}{b^2|\bm{x}_{\sss\rm 0}|^3}\!-\!\frac{6(\bm{n}\!\cdot\!\bm{J})(\bm{n}\!\cdot\!\bm{x}_{\sss\rm 0})}{|\bm{x}_{\sss\rm 0}|^5}\Big\},
\end{eqnarray}}
{\small \begin{eqnarray}
 && \hskip -2.cm \frac{d\bm{x}}{dt}
\!=\!v_e\bm{n}\,\Big\{1\!-\!\Big(\!3\!-\!\frac{1}{v_e^2}\!\Big)\!\Big(\!1\!+\!\frac{2m}{v_e^2|\bm{x}_{\sss\rm e}|}\Big)\!\Big(\!\frac{m}{|\bm{x}_{\sss\rm 0}|}\!-\!\frac{m}{|\bm{x}_{\sss\rm e}|}\!\Big)\!
-\!\Big(\!9\!-\!\frac{6}{v_e^2}\!+\!\frac{1}{v_e^4}\!\Big)\!\frac{m^2}{|\bm{x}_{\sss\rm 0}|^2}\ln\!\frac{|\bm{x}_{\sss\rm 0}|}{|\bm{x}_{\sss\rm e}|}
\!\Big.\nn\\
 && \hskip -0.2cm \Big.-\!\Big[\!\Big(\frac{9}{2}\!-\!\frac{4}{v_e^2}\!+\!\frac{3}{2v_e^4}\!\Big)\!-\!\Big(\frac{3}{2}\!
 -\!\frac{1}{2v_e^2}\!\Big)\!\frac{q^2}{m^2}\Big]\!\Big(\!\frac{m^2}{|\bm{x}_{\sss\rm 0}|^2}\!-\!\frac{m^2}{|\bm{x}_{\sss\rm e}|^2}\!\Big)
 \Big\}\nn\\
  &&  \hskip -1.45cm  +v_e\bm{b}\,\Big\{
\!\Big[\Big(\frac{1}{2}\!+\!\frac{1}{2v_e^2}\Big)\!+\!\Big(\frac{5}{2}\!-\!\frac{5}{v_e^2}\!+\!\frac{1}{2v_e^4}\!\Big)
\!\frac{m}{|\bm{x}_{\sss\rm e}|}\!+\!\frac{(\bm{n}\!\times\!\bm{b})\!\cdot\!\bm{J}}{v_eb^2m}\Big]\!\Big(\!\frac{m}{|\bm{x}_{\sss\rm 0}|^2}\!-\!\frac{m}{|\bm{x}_{\sss\rm e}|^2}\!\Big)\!\Big.\nn\\
&&  \hskip -0.2cm \Big.
+\!\Big[\Big(\frac{3}{2}\!+\!\frac{1}{v_e^2}\!-\!\frac{1}{2v_e^4}\!\Big)\!\frac{m^2}{|\bm{x}_{\sss\rm e}|^2}\!-\!\Big(\!1\!-\!\frac{4}{v_e^2}\!+\!\frac{3}{v_e^4}\!\Big)\!\frac{m^2}{|\bm{x}_{\sss\rm 0}|^2}\Big]\!\Big(\!\frac{1}{|\bm{x}_{\sss\rm 0}|}\!-\!\frac{1}{|\bm{x}_{\sss\rm e}|}\!\Big)\Big.\nn\\
&&  \hskip -0.2cm\Big.+\!\Big(\frac{6}{v_e^2}\!-\!\frac{2}{v_e^4}\!\Big)\!\frac{m^2}{|\bm{x}_{\sss\rm 0}|^3}\ln\!\frac{|\bm{x}_{\sss\rm 0}|}{|\bm{x}_{\sss\rm e}|}\!
-\!\Big[\!\Big(\frac{5}{2}\!-\!\frac{7}{3v_e^2}\!-\!\frac{1}{6v_e^4}\!\Big)\!+\!\Big(\frac{1}{3}\!
-\!\frac{1}{3v_e^2}\!\Big)\!\frac{q^2}{m^2}\Big]\Big(\!\frac{m^2}{|\bm{x}_{\sss\rm 0}|^3}\!-\!\frac{m^2}{|\bm{x}_{\sss\rm e}|^3}\!\Big)\Big\}\nn\\
&&  \hskip -1.65cm+ \!\bm{n}\!\times\!\bm{b} \,\Big\{2(\bm{n}\!\cdot\!\bm{J})\Big(\!\frac{1}{|\bm{x}_{\sss\rm 0}|^3}\!-\!\frac{1}{|\bm{x}_{\sss\rm e}|^3}\!\Big)-\frac{\bm{b}\!\cdot\!\bm{J}}{b^2}\!\Big(\!\frac{1}{|\bm{x}_{\sss\rm 0}|^2}\!-\!\frac{1}{|\bm{x}_{\sss\rm e}|^2}\!\Big)\Big\}~,
\end{eqnarray}}
{\small \begin{eqnarray}
 && \hskip -2cm \bm{x}
\!= \!|\bm{x}_{\sss\rm 0}| \bm{n}\,\Big\{\!1\!
-\!\frac{m}{|\bm{x}_{\sss\rm 0}|}\!\Big(\!3\!-\!\frac{1}{v_e^2}\!\Big)\!\Big(\!\ln\!\frac{|\bm{x}_{\sss\rm 0}|}{|\bm{x}_{\sss\rm e}|}\!-\!\frac{|\bm{x}_{\sss\rm 0}|}{|\bm{x}_{\sss\rm e}|}\!+\!1\!\Big)\!
\!+\!\frac{m^2}{|\bm{x}_{\sss\rm 0}|^2}\!\Big[\!\Big(\!9\!-\!\frac{6}{v_e^2}\!+\!\frac{1}{v_e^4}\!\Big)\!-\!
\!\Big(\!\frac{6}{v_e^2}\!-\!\frac{2}{v_e^4}\!\Big)\!\frac{|\bm{x}_{\sss\rm 0}|}{|\bm{x}_{\sss\rm e}|}
\Big]\!\ln\!\frac{|\bm{x}_{\sss\rm 0}|}{|\bm{x}_{\sss\rm e}|}\!\nn\\
&& \hskip -1.6cm \!+\!\frac{m^2}{|\bm{x}_{\sss\rm 0}||\bm{x}_{\sss\rm e}|}\!\Big[\!\Big(\frac{9}{2}\!+\!\frac{2}{v_e^2}\!-\!\frac{1}{2v_e^4}\!\Big)\!\Big(\!\frac{|\bm{x}_{\sss\rm 0}|}{|\bm{x}_{\sss\rm e}|}\!-\!1\!\Big)\!+\!\Big(\frac{27}{2}\!-\!\frac{10}{v_e^2}\!+\!\frac{5}{2v_e^4}\!\Big)
\!\Big(\!\frac{|\bm{x}_{\sss\rm e}|}{|\bm{x}_{\sss\rm 0}|}\!-\!1\!\Big)\!\Big] \!-\!\frac{3v_e^2\!-\!1}{2v_e^2}\!\Big(\!\frac{q}{|\bm{x}_{\sss\rm 0}|}\!-\!\frac{q}{|\bm{x}_{\sss\rm e}|}\!\Big)^{\!2}\Big\}\nn \\
&& \hskip -1.8cm +\bm{b}\,\Big\{\!\!-\!\Big(\frac{1}{2}\!+\!\frac{1}{2v_e^2}\Big)\!\frac{m}{|\bm{x}_{\sss\rm 0}|}\!\Big(\!1\!-\!\frac{|\bm{x}_{\sss\rm 0}|}{|\bm{x}_{\sss\rm e}|}\Big)^{\!2\!}\!-\!\Big[\Big(\!\frac{3}{v_e^2}\!-\!\frac{1}{v_e^4}\Big)\!\frac{m^2}{|\bm{x}_{\sss\rm 0}|^2}\!-\!\Big(\frac{3}{2}\!+\!\frac{1}{v_e^2}\!-\!\frac{1}{2v_e^4}\Big)\!\frac{m^2}{|\bm{x}_{\sss\rm e}|^2}\Big]\ln\!\frac{|\bm{x}_{\sss\rm 0}|}{|\bm{x}_{\sss\rm e}|}\nn\\
&&\hskip -1.cm \!+\!\Big[\!\Big(\frac{3}{2}\!-\!\frac{5}{3v_e^2}\!+\!\frac{1}{6v_e^4}\!\Big)\!\frac{m^2}{|\bm{x}_{\sss\rm e}|^2}\!-\!\Big(\frac{7}{2}\!-\!\frac{9}{v_e^2}\!+\!\frac{7}{2v_e^4}\!\Big)\!\frac{m^2}{|\bm{x}_{\sss\rm e}||\bm{x}_{\sss\rm 0}|}\!+\!\Big(\frac{1}{3}\!-\!\frac{1}{3v_e^2}\!\Big)\!\frac{q^2}{|\bm{x}_{\sss\rm e}|^2}\Big]\!\Big(\!1\!-\!\frac{|\bm{x}_{\sss\rm 0}|}{|\bm{x}_{\sss\rm e}|}\!\Big)\!\nn\\
&&\hskip -1cm \!+\!\Big[\!\Big(\frac{7}{4}\!-\!\frac{14}{3v_e^2}\!+\!\frac{23}{12v_e^4}\Big)\!-\!\Big(\frac{1}{6}\!-\!\frac{1}{6v_e^2}\!\Big)\!
\frac{q^2}{m^2}\Big]\!\Big(\!\frac{m^2}{|\bm{x}_{\sss\rm 0}|^2}\!-\!\frac{m^2}{|\bm{x}_{\sss\rm e}|^2}\!\Big)\!-\!\frac{(\bm{n}\!\times\!\bm{b})\!\cdot\!\bm{J}}{v_eb^2|\bm{x}_{\sss\rm 0}|}\!\Big(\!1\!-\!\frac{|\bm{x}_{\sss\rm 0}|}{|\bm{x}_{\sss\rm e}|}\!\Big)^{\!2}\Big\}\nn\\
&& \hskip -1.5cm
+\bm{n}\!\times\!\bm{b}\,\Big\{\!-\!\frac{\bm{n}\!\cdot\!\bm{J}}{v_e}\!\Big[\!\Big(\!\frac{1}{|\bm{x}_{\sss\rm 0}|^2}\!-\!\frac{1}{|\bm{x}_{\sss\rm e}|^2}\!\Big)\!+\!\frac{2(|\bm{x}_{\sss\rm 0}|\!-\!|\bm{x}_{\sss\rm e}|)}{|\bm{x}_{\sss\rm e}|^3}\Big]\!+\!\frac{\bm{b}\!\cdot\!\bm{J}}{v_eb^2|\bm{x}_{\sss\rm 0}|}\!\Big(\!1\!-\!\frac{|\bm{x}_{\sss\rm 0}|}{|\bm{x}_{\sss\rm e}|}\!\Big)^{\!2}
\Big\}~,
\end{eqnarray}}
\noindent \hskip -0.1cm here we have dropped the terms whose magnitudes are proportional to or higher orders than $b^2$.

\section{Two special cases}\label{sec:4th}

In this section we consider two special scenarios. For all cases, the Minkowskian solutions to the velocity $\frac{d\bm{x}_{\sss\rm 0}}{dt}$ and the trajectory $\bm{x}_{\sss\rm 0}$ are described by Eqs.\,(\ref{NewtonianMotionV}) and (\ref{NewtonianMotionT}), so we only list the 1PM and 2PM contributions to the velocity and the trajectory in the below. It is worth emphasizing that the 1PM contributions consist of the Newtonian-term contribution and the first-order relativistic contribution.

\subsection{Case 1: the particle passes by the black hole with
$|\bm{x}_{\rm e}|\!\rightarrow\!-\infty,~|\bm{x}_{\sss\rm 0}|\!\rightarrow\!+\infty,~v_e \gg m/b$}

In this case, the PM corrections to the Minkowskian solutions of the test particle's velocity and trajectory can be written as follows
{\small \begin{eqnarray}\label{eq:velocity-infty}
&& \hskip -1cm \frac{d\bm{x}_{\sss\rm 1PM}}{dt} = v_e\frac{\bm{b}}{b} \Big[\!-\!\Big(\!1\!+\frac{1}{v_e^2}\!\Big)\frac{2m}{b} \Big]~,
\end{eqnarray}}
{\small \begin{eqnarray}\label{eq:velocity-infty}
&&  \hskip -1cm \frac{d\bm{x}_{\sss\rm 2PM}}{dt} = v_e\bm{n}\Big[\! -\!\Big(\!1\!+\frac{2}{v_e^2}\!+\frac{1}{v_e^4}\!\Big)\!\frac{2m^2}{b^2}\Big]
+v_e\bm{n}\!\times\!\frac{\bm{b}}{b}\Big(\!\!-\!\frac{4\bm{b}\!\cdot\!\bm{J}}{v_eb^3}\Big)\nn\\
&&  \hskip 1.35cm   + v_e\frac{\bm{b}}{b} \Big[\! \Big(\!\frac{3}{4}\!-\!\frac{5}{v_e^2}\!+\!\frac{1}{2v_e^4}\!\Big)\!\frac{\pi m^2}{b^2} \!+\! \Big(\!\frac{1}{4}\!+\!\frac{1}{2v_e^2}\!\Big)\!\frac{\pi q^2}{b^2} \!-\! \frac{4(\bm{n}\times\bm{b})\!\cdot\!\bm{J}}{v_eb^{3}}\Big]~,
\end{eqnarray}}
{\small \begin{equation}
\hskip -1cm \bm{x}_{\sss\rm 1PM} = m\bm{n}\Big[\!\Big(3\!-\!\frac{1}{v_e^2}\Big)\!\frac{\bm{n}\!\cdot\!(\bm{x}_{\sss\rm 0}\!-\!\bm{x}_{\sss\rm e})}{|\bm{x}_{\sss\rm e}|}\Big] - m\frac{\bm{b}}{b}\Big[\!\Big(\!1\!+\!\frac{1}{v_e^2}\Big)\!\Big(\!1\!-\!\frac{\bm{x}_{\sss \rm e}\!\cdot\!\bm{x}_{\sss\rm 0}}{|\bm{x}_{\sss \rm e}||\bm{x}_{\sss\rm 0}|}\!\Big)\!\frac{|\bm{x}_{\sss\rm 0}|}{b}\Big]~,\label{eq:trajectory-2inf-1pn}\end{equation}}
{\small \begin{eqnarray}
&& \hskip -1cm \bm{x}_{\sss\rm 2PM} = \frac{m^2\bm{n}}{b}\Big\{\!\!-\!\Big(\!2\!+\!\frac{4}{v_e^2}\!+\!\frac{2}{v_e^4}\!\Big)\!
\frac{|\bm{x}_{\sss\rm 0}|}{b} \Big\}\!-\bm{n}\!\times\!\bm{b}\,\Big\{\frac{2|\bm{x}_{\sss\rm 0}|\bm{b}\!\cdot\!\bm{J}}{v_eb^4}\!\Big(\!1\!-\!\frac{\bm{x}_{\sss \rm e}\!\cdot\!\bm{x}_{\sss\rm 0}}{|\bm{x}_{\sss \rm e}||\bm{x}_{\sss\rm 0}|}\!\Big)\Big\}\!\nn\\
&& \hskip -.08cm  +\frac{m^2\bm{b}}{b^2}\Big\{\!\!-\!\Big(\!5\!-\frac{10}{v_e^2}\!+\!\frac{1}{v_e^4}\!\Big)\!\frac{|\bm{x}_{\sss\rm 0}|}{|\bm{x}_{\sss \rm e}|}
\!+\!\Big(\frac{3}{4}\!-\!\frac{5}{v_e^2}\!+\!\frac{1}{2v_e^4}\!\Big)\frac{\pi|\bm{x}_{\sss\rm 0}|\!}{b}
+\Big(\frac{17}{4}\!-\!\frac{5}{v_e^2}\!+\!\frac{1}{2v_e^4}\!\Big)\!\frac{\bm{x}_{\sss \rm e}\!\cdot\!\bm{x}_{\sss\rm 0}}{|\bm{x}_{\sss \rm e}|^2} \nn\\
&& \hskip 1.4cm -\!\Big(\frac{1}{4}\!+\!\frac{1}{2v_e^2}\!\Big)\!\frac{q^2}{m^2}\!\Big(\!\frac{\bm{x}_{\sss \rm e}\!\cdot\!\bm{x}_{\sss\rm 0}}{|\bm{x}_{\sss \rm e}|^2}\!-\!\frac{\!\pi|\bm{x}_{\sss\rm 0}|\!}{b}\Big)\!-\!\frac{2|\bm{x}_{\sss\rm 0}|(\bm{n}\!\times\!\bm{b})\!\cdot\!\bm{J}}{v_e\,b^2\,m^2}\!\Big(\!1\!-\!\frac{\bm{x}_{\sss \rm e}\!\cdot\!\bm{x}_{\sss\rm 0}}{|\bm{x}_{\sss \rm e}||\bm{x}_{\sss\rm 0}|}\!\Big)\Big\}~.\label{eq:trajectory-2inf-2PN}
\end{eqnarray}}
\noindent \hskip -0.2cm For simplicity, Eqs.\,(\ref{eq:trajectory-2inf-1pn}) and (\ref{eq:trajectory-2inf-2PN}) only keep the leading terms whose magnitudes are proportional to $|\bm{x}_{\sss\rm 0}|$.

\subsection{Case 2: the particle leaves away from some position near to the black hole with $v_e \!\gg\! m/|\bm{x}_{\sss\rm e}|$ and $|\bm{x}_{\sss\rm 0}|\!\rightarrow\!+\infty$}
In this case we can simplify the PM contributions to the velocity and the trajectory of the test particle as follows
{\small \begin{eqnarray}\label{eq:velocity-infty}
&&  \hskip -2cm \frac{d\bm{x}_{\sss\rm 1PM}}{dt} = v_e\bm{n}\Big[\!\Big(3\!-\frac{1}{v_e^2}\!\Big)\!\frac{m}{|\bm{x}_{\sss \rm e}|}\Big]+ v_e\frac{\bm{b}}{b} \Big[\!-\!\Big(\!1\!+\frac{1}{v_e^2}\!\Big)\!\frac{m}{b}\!\Big(\!1\!-\!\frac{\bm{n}\!\cdot\!\bm{x}_{\sss\rm e}}{|\bm{x}_{\sss\rm e}|}\Big) \!\Big]~,
\end{eqnarray}}
{\small \begin{eqnarray}
 && \hskip -2cm \frac{d\bm{x}_{\sss\rm 2PM}}{dt}
\!=\!v_e \bm{n}\Big\{\frac{m^2}{|\bm{x}_{\sss\rm e}|^2}\!\Big[\!\Big(\!5\!+\!\frac{3}{v_e^2}\!\Big)\!-
\!\Big(\frac{3}{2}\!-\!\frac{1}{2v_e^2}\!\Big)\!\frac{q^2}{m^2}\!-\!\frac{(m^2\!-\!q^2)b^2}{2m^2|\bm{x}_{\sss\rm e}|^2}\Big]\!-\!\Big(\!1\!+\!\frac{2}{v_e^2}\!+\!\frac{1}{v_e^4}\!\Big)
\!\Big(\!1\!-\!\frac{\bm{n}\!\cdot\!\bm{x}_{\sss\rm e}}{|\bm{x}_{\sss\rm e}|}\!\Big)\!\frac{m^2}{b^2}\Big\}\nn\\
&& \hskip -1.cm +v_e\frac{\bm{b}}{b}\Big\{\Big[\!\Big(\frac{17}{4}\!-\!\frac{5}{v_e^2}\!+\!\frac{1}{2v_e^4}\Big)\!\frac{m^2}{b}\!
 -\!\Big(\frac{1}{4}\!+\!\frac{1}{2v_e^2}\Big)\!\frac{q^2}
{b}\!+\!\frac{(m^2\!-\!q^2)b}{2|\bm{x}_{\sss\rm e}|^2}\Big]\!\frac{\bm{n}\!\cdot\!\bm{x}_{\sss\rm e}}{|\bm{x}_{\sss\rm e}|^2}\!-\!\Big(\!5\!-\!\frac{10}{v_e^2}\!+\!\frac{1}{v_e^4}\Big)\!\frac{m^2}{b|\bm{x}_{\sss\rm e}|}\nn\\
 &&  \hskip 0.3cm +\!\Big[\!\Big(\frac{3}{4}\!-\!\frac{5}{v_e^2}\!+\!\frac{1}{2v_e^4}\!\Big)\!\frac{m^2}{b^2}\!
 +\!\Big(\frac{1}{4}\!+\!\frac{1}{2v_e^2}\!\Big)
\!\frac{q^2}{b^2}\Big]\!\arccos\frac{\bm{n}\!\cdot\!\bm{x}_{\sss\rm e}}{|\bm{x}_{\sss\rm e}|}\!-\!\frac{2(\bm{n}\!\times\!\bm{b})\!\cdot\!\bm{J}}{v_eb^3}\!\Big(\!1\!-\!\frac{\bm{n}\!\cdot\!\bm{x}_{\sss\rm e}}{|\bm{x}_{\sss\rm e}|}\!\Big)\!\Big\}\nn\\
&&  \hskip -1.3cm +\bm{n}\!\times\!\frac{\bm{b}}{b}\Big\{\!-\!\frac{2b\bm{n}\!\cdot\!\bm{J}}{|\bm{x}_{\sss\rm e}|^3}
-\frac{2\bm{b}\!\cdot\!\bm{J}}{b}\!\Big(\frac{1}{b^2}\!-\!\frac{\bm{n}\!\cdot\!\bm{x}_{\sss\rm e}}{b^2|\bm{x}_{\sss\rm e}|}\!-\!
\frac{\bm{n}\!\cdot\!\bm{x}_{\sss\rm e}}{|\bm{x}_{\sss\rm e}|^3}\Big)\Big\},
\end{eqnarray}}
{\small \begin{eqnarray}
&& \hskip -2cm \bm{x}_{\sss\rm 1PM} =
m\bm{n}\Big[\!\Big(\!3\!-\!\frac{1}{v_e^2}\Big)\!\frac{\bm{n}\!\cdot\!\bm{x}_{\sss\rm 0}}{|\bm{x}_{\sss\rm e}|}\Big] +m\frac{\bm{b}}{b}\Big[\!-\!\Big(\!1\!+\!\frac{1}{v_e^2}\Big)\!\Big(\!1\!-\!\frac{\bm{x}_{\sss \rm e}\!\cdot\!\bm{x}_{\sss\rm 0}}{|\bm{x}_{\sss \rm e}||\bm{x}_{\sss\rm 0}|}\!\Big)\frac{|\bm{x}_{\sss\rm 0}|}{b}\Big]~,\label{eq:trajectory-particle-1RC-1inf}
\end{eqnarray}}
{\small \begin{eqnarray}
 &&\hskip -2cm \bm{x}_{\sss\rm 2PM}
\!=\!\frac{m^2}{b}\bm{n}\Big\{\!\Big[\!\Big(\!5\!+\!\frac{3}{v_e^2}\!\Big)\!-\!\frac{(m^2\!-\!q^2)b^2}{2|\bm{x}_{\sss\rm e}|^2}\!
-\!\Big(\frac{3}{2}\!
-\!\frac{1}{2v_e^2}\!\Big)\!\frac{q^2}{m^2}\!\Big]\!\frac{b|\bm{x}_{\sss\rm 0}|}{|\bm{x}_{\sss\rm e}|^2}\!-\!\Big(\!1\!+\!\frac{2}{v_e^2}\!+\!\frac{1}{v_e^4}\!\Big)\!\Big(\!1\!-\!\frac{\bm{n}\!\cdot\!\bm{x}_{\sss\rm e}}{|\bm{x}_{\sss\rm e}|}\!\Big)\!\frac{|\bm{x}_{\sss\rm 0}|}{b}\!\Big\}\nn\\
 &&\hskip -1.15cm +\!\frac{m^2\bm{b}}{b^2}\Big\{\!\Big[\!\Big(\frac{17}{4}\!-\!\frac{5}{v_e^2}\!+\!\frac{1}{2v_e^4}\!\Big)\!
-\!\Big(\frac{1}{4}\!+\!\frac{1}{2v_e^2}\!\Big)\!\frac{q^2}{m^2}\!+\!\frac{(m^2-q^2)b^2}{2m^2|\bm{x}_{\sss\rm e}|^2}\Big]\!\frac{\bm{x}_{\sss\rm e}\!\cdot\!\bm{x}_{\sss\rm 0}}{|\bm{x}_{\sss\rm e}|^2}\!-\!\Big(\!5\!-\!\frac{10}{v_e^2}\!+\!\frac{1}{v_e^4}\!\Big)
\!\frac{|\bm{x}_{\sss\rm 0}|}{|\bm{x}_{\sss\rm e}|}\!\nn\\
&& \hskip -0.8cm +\!\frac{|\bm{x}_{\sss\rm 0}|}{b}\!\Big[\!\Big(\frac{3}{4}\!-\!\frac{5}{v_e^2}\!+\!\frac{1}{2v_e^4}\!\Big)\!+\!\Big(\frac{1}{4}\!+\!\frac{1}{2v_e^2}\!\Big)
\!\frac{q^2}{m^2}\Big]\!\Big(\!\arccos\frac{\bm{n}\!\cdot\!\bm{x}_{\sss\rm e}}{|\bm{x}_{\sss\rm e}|}\!\Big)\!-\!\frac{2|\bm{x}_{\sss\rm 0}|(\bm{n}\!\times\!\bm{b})\!\cdot\!\bm{J}}{v_em^2b^2}\!\Big(\!1\!-\!\frac{\!\bm{x}_{\sss\rm e}\!\cdot\!\bm{x}_{\sss\rm 0}}{\!|\bm{x}_{\sss\rm e}||\bm{x}_{\sss\rm 0}|}\!\Big)\!\Big\}\nn\\
&& \hskip -1.15cm
+\bm{n}\!\times\!\bm{b}\Big\{\!\!-\!\frac{2\bm{n}\!\cdot\!\bm{J}}{v_e}\frac{|\bm{x}_{\sss\rm 0}|}{|\bm{x}_{\sss\rm e}|^3}-\frac{2\bm{b}\!\cdot\!\bm{J}}{b^2v_e}\!\Big[\frac{|\bm{x}_{\sss\rm 0}|}{b^2}\!-\!\frac{\bm{x}_{\sss\rm e}\!\cdot\!\bm{x}_{\sss\rm 0}}{|\bm{x}_{\sss\rm e}|}\!\Big(\frac{1}{b^2}\!+\!\frac{1}{|\bm{x}_{\sss\rm e}|^2}\!\Big)\Big]\Big\}~.
\label{eq:trajectory-particle-2RC-1inf}
\end{eqnarray}}
\hskip -0.1cm Again, Eqs.\,(\ref{eq:trajectory-particle-1RC-1inf}) and (\ref{eq:trajectory-particle-2RC-1inf}) only keep the leading terms whose magnitudes are proportional to $|\bm{x}_{\sss\rm 0}|$.


%
%

\section{Discussions and Summaries}\label{sec:5th}
In this work we have derived the 2PM analytical solutions to the velocity and the trajectory of the small-deflection particle in the Kerr-Newman spacetime, via an iterative method which can be used for any spacetime and up to an arbitrary PM order in principle. The particle's trajectory is described by the combination of Eqs.\,(\ref{eq:trajectory-2RC-general}),\,(\ref{NewtonianMotionT}),\,(\ref{eq:trajectory-particle-N}),\,
(\ref{eq:trajectory-particle-1RC}),\,(\ref{2RCTrajectory}),\,(\ref{2RCTrajectory-para}) and (\ref{2RCTrajectory-perp}). The velocity is described by the combination of Eqs.\,(\ref{eq:velocity-2RC-general}),\,(\ref{NewtonianMotionV}),\,(\ref{eq:velocity-particle-N}),\,(\ref{eq:velocity-particle-1RC}) and (\ref{eq:velocity-2PM})-(\ref{eq:velocity-2PM-perp}). In these formulas, all the arguments of the natural logarithm are positive, and there is no issue about choosing its branches. All the arguments of the trigonometric function $\arccos$ are in the range of [-1, 1], and only its principal branch is adopted. It is worth emphasizing that these solutions are applicable only under the small-deflection condition $v_e^2\!\gg\!m/|\bm{x}_{\sss\rm 0}|$ being satisfied.

We have checked that these formulas recover the 2PN solutions of the photon's trajectory and velocity in the Kerr-Newman spacetime~\cite{JiangLin2018}, when the initial velocity of the test particle is set as
{\small \begin{equation}
v_e=1-\frac{2m}{|\bm{x}_{\sss\rm e}|}+\frac{5m^2\!+\!q^2}{2\bm{x}_{\sss\rm e}^2}-\frac{(m^2\!-\!q^2)(\bm{n}\!\cdot\!\bm{x}_{\sss\rm e})^2}{2\bm{x}_{\sss\rm e}^4}-\frac{2 \bm{n}\!\cdot\!(\bm{x}\!\times\!\bm{J})}{|\bm{x}_{\sss\rm e}|^3}~,\label{LightV}
\end{equation}}
\noindent \hskip -0.05cm which is prescribed by the null-geodesic condition
to the 2PM accuracy. The derivation of Eq.\,(\ref{LightV}) can be found in \ref{Derivation4LightV}. 
Therefore, the obtained formulas provide an unified description for the small-deflection motion of the photon and the massive particle in the Kerr-Newman spacetime to the 2PM accuracy. When the source's angular momentum and charge are dropped, our solutions reduce to the 1PM and 2PM solutions of the photon's motion in the Schwarzschild spacetime~\cite{Will1981,Klioner2010}. 
Under the small-deflection approximation, this work also extends the ones about the massive particle's motion in the Schwarzschild spacetime~\cite{AcciolyRagusa2002,BhadraSarkarNandi2007}, and the gravitational deflection of the massive particle in the equatorial plane of the Kerr-Newman spacetime~\cite{HeLin2017}.

Finally, from the applied views, these formulas can be used to predict the trajectory and the velocity of a body passing by a rotating black hole with a large impact factor, or the motion of a body moving inwards/outwards the black hole with any impact factor until the small-deflection condition becomes invalid. Since the analytical solutions can be evaluated directly without numerically solving the geodesic equation, they might be integrated into the geodesic-equation solver in the theoretical templates to speed up the calculations about the stellar's motion in the strong gravitational fields. 

\appendix

\section{Lists of integrals}\label{Integrals}
The key integrals used in the derivations are listed as follows
{\small
\begin{eqnarray*}
&& \hskip -2cm \int_{t_{\sss \rm e}}^t \!v_e dt \!=\!\bm{n}\!\cdot\! \bm{x}_{\sss\rm 0} \!-\!\bm{n}\!\cdot\! \bm{x}_{\sss \rm e}, \\
&& \hskip -2cm \int_{t_{\sss \rm e}}^t \!\frac{v_e}{|\bm{x}_{\sss\rm 0}|} dt \!=\! \ln\!{\frac{|\bm{x}_{\sss\rm 0}|\!+\!\bm{n}\!\cdot\!\bm{x}_{\sss\rm 0}}{|\bm{x}_{\sss \rm e}|\!+\!\bm{n}\!\cdot\!\bm{x}_{\sss \rm e}}},\\
&& \hskip -2cm \int_{t_{\sss \rm e}}^t \!v_e\frac{\bm{n}\!\cdot\!\bm{x}_{\sss\rm 0}}{|\bm{x}_{\sss\rm 0}|} dt \!=\! |\bm{x}_{\sss\rm 0}| \!-\! |\bm{x}_{\sss \rm e}|, \\
&& \hskip -2cm \int_{t_{\sss \rm e}}^t \!\frac{v_e}{|\bm{x}_{\sss\rm 0}|^2} dt \!=\!-\frac{1}{b}\!\Big(\!\arccos\!{\frac{\bm{n}\!\cdot\!\bm{x}_{\sss\rm 0}}{|\bm{x}_{\sss\rm 0}|}} \!-\! \arccos\!{\frac{\bm{n}\!\cdot\!\bm{x}_{\sss \rm e}}{|\bm{x}_{\sss \rm e}|}}\!\Big), \\
&& \hskip -2cm \int_{t_{\sss \rm e}}^t \!v_e\frac{\bm{n}\!\cdot\!\bm{x}_{\sss\rm 0}}{|\bm{x}_{\sss\rm 0}|^2} dt \!=\! \ln\!{\frac{|\bm{x}_{\sss\rm 0}|}{|\bm{x}_{\sss \rm e}|}}, \\
&& \hskip -2cm \int_{t_{\sss \rm e}}^t \!\frac{v_e}{|\bm{x}_{\sss\rm 0}|^3} dt \!=\! \frac{1}{b^2}\!\Big(\!\frac{\bm{n}\!\cdot\!\bm{x}_{\sss\rm 0}}{|\bm{x}_{\sss\rm 0}|} \!-\! \frac{\bm{n}\!\cdot\!\bm{x}_{\sss \rm e}}{|\bm{x}_{\sss \rm e}|}\!\Big), \\
&& \hskip -2cm \int_{t_{\sss \rm e}}^t \!v_e\frac{\bm{n}\!\cdot\!\bm{x}_{\sss\rm 0}}{|\bm{x}_{\sss\rm 0}|^3} dt \!=\! -\frac{1}{|\bm{x}_{\sss\rm 0}|} \!+\! \frac{1}{|\bm{x}_{\sss \rm e}|}, \\
&& \hskip -2cm \int_{t_{\sss \rm e}}^t \!\frac{v_e}{|\bm{x}_{\sss\rm 0}|^4} dt \!=\! \frac{1}{2b^2}\!\Big(\!\frac{\bm{n}\!\cdot\!\bm{x}_{\sss\rm 0}}{|\bm{x}_{\sss\rm 0}|^2} \!-\! \frac{\bm{n}\!\cdot\!\bm{x}_{\sss \rm e}}{|\bm{x}_{\sss \rm e}|^2}\!\Big)\!-\!\frac{1}{2b^3}\!\Big(\!\arccos\!{\frac{\bm{n}\!\cdot\!\bm{x}_{\sss\rm 0}}{|\bm{x}_{\sss\rm 0}|}} \!-\! \arccos\!{\frac{\bm{n}\!\cdot\!\bm{x}_{\sss \rm e}}{|\bm{x}_{\sss \rm e}|}}\!\Big), \\
&& \hskip -2cm \int_{t_{\sss \rm e}}^t \!v_e\frac{\bm{n}\!\cdot\!\bm{x}_{\sss\rm 0}}{|\bm{x}_{\sss\rm 0}|^4} dt \!=\! -\frac{1}{2|\bm{x}_{\sss\rm 0}|^2} \!+\! \frac{1}{2|\bm{x}_{\sss \rm e}|^2}, \\
&& \hskip -2cm \int_{t_{\sss \rm e}}^t \!\frac{v_e}{|\bm{x}_{\sss\rm 0}|^5} dt \!=\! \frac{1}{3b^2}\!\Big(\!\frac{\bm{n}\!\cdot\!\bm{x}_{\sss\rm 0}}{|\bm{x}_{\sss\rm 0}|^3} \!-\! \frac{\bm{n}\!\cdot\!\bm{x}_{\sss \rm e}}{|\bm{x}_{\sss \rm e}|^3}\!\Big) \!+\! \frac{2}{3b^4}\!\Big(\!\frac{\bm{n}\!\cdot\!\bm{x}_{\sss\rm 0}}{|\bm{x}_{\sss\rm 0}|} \!-\! \frac{\bm{n}\!\cdot\!\bm{x}_{\sss \rm e}}{|\bm{x}_{\sss \rm e}|}\!\Big), \\
&& \hskip -2cm \int_{t_{\sss \rm e}}^t \!v_e\frac{\bm{n}\!\cdot\!\bm{x}_{\sss\rm 0}}{|\bm{x}_{\sss\rm 0}|^5} dt \!=\! -\frac{1}{3|\bm{x}_{\sss\rm 0}|^3} \!+\! \frac{1}{3|\bm{x}_{\sss \rm e}|^3}, \\
&& \hskip -2cm \int_{t_{\sss \rm e}}^t \!\frac{v_e}{|\bm{x}_{\sss\rm 0}|^6} dt \!=\! \frac{3}{8b^4}\!\Big(\!\frac{\bm{n}\!\cdot\!\bm{x}_{\sss\rm 0}}{|\bm{x}_{\sss\rm 0}|^2} \!-\! \frac{\bm{n}\!\cdot\!\bm{x}_{\sss \rm e}}{|\bm{x}_{\sss \rm e}|^2}\!\Big) \!+\! \frac{1}{4b^2}\!\Big(\!\frac{\bm{n}\!\cdot\!\bm{x}_{\sss\rm 0}}{|\bm{x}_{\sss\rm 0}|^4} \!-\! \frac{\bm{n}\!\cdot\!\bm{x}_{\sss \rm e}}{|\bm{x}_{\sss \rm e}|^4}\!\Big)\!-\!\frac{3}{8b^5}\!\Big(\!\arccos\!{\frac{\bm{n}\!\cdot\!\bm{x}_{\sss\rm 0}}{|\bm{x}_{\sss\rm 0}|}} \!-\! \arccos\!{\frac{\bm{n}\!\cdot\!\bm{x}_{\sss \rm e}}{|\bm{x}_{\sss \rm e}|}}\!\Big), \\
&& \hskip -2cm \int_{t_{\sss \rm e}}^t \!v_e\frac{\bm{n}\!\cdot\!\bm{x}_{\sss\rm 0}}{|\bm{x}_{\sss\rm 0}|^6} dt \!=\! -\frac{1}{4|\bm{x}_{\sss\rm 0}|^4} \!+\! \frac{1}{4|\bm{x}_{\sss \rm e}|^4}, \\
&& \hskip -2cm \int_{t_{\sss \rm e}}^t \!v_e\arccos\!{\frac{\bm{n}\!\cdot\!\bm{x}_{\sss\rm 0}}{|\bm{x}_
{\sss\rm 0}|}} dt \!=\! \Big[(\bm{n}\!\cdot\!\bm{x}_{\sss\rm 0})\!\arccos\!{\frac{\bm{n}\!\cdot\!\bm{x}_{\sss\rm 0}}{|\bm{x}_{\sss\rm 0}|}} \!-\! (\bm{n}\!\cdot\!\bm{x}_{\sss \rm e})\!\arccos\!{\frac{\bm{n}\!\cdot\!\bm{x}_{\sss \rm e}}{|\bm{x}_{\sss \rm e}|}} \Big]\!+\! b \ln\!{\frac{|\bm{x}_{\sss\rm 0}|}{|\bm{x}_{\sss \rm e}|}},\\
&& \hskip -2cm \int_{t_{\sss \rm e}}^t \!\frac{v_e}{|\bm{x}_{\sss\rm 0}|^3} \!\ln\!{\frac{|\bm{x}_{\sss\rm 0}|\!+\!\bm{n}\!\cdot\!\bm{x}_{\sss\rm 0}}{|\bm{x}_{\sss \rm e}|\!+\!\bm{n}\!\cdot\!\bm{x}_{\sss \rm e}}} dt\!=\! \frac{1}{b^2}\frac{\bm{n}\!\cdot\!\bm{x}_{\sss\rm 0}}{|\bm{x}_{\sss\rm 0}|}\!\ln\!{\frac{|\bm{x}_{\sss\rm 0}|\!+\!\bm{n}\!\cdot\!\bm{x}_{\sss\rm 0}}{|\bm{x}_{\sss \rm e}|\!+\!\bm{n}\!\cdot\!\bm{x}_{\sss \rm e}}}\!-\! \frac{1}{b^2}\!\ln\!{\frac{|\bm{x}_{\sss\rm 0}|}{|\bm{x}_{\sss \rm e}|}},\\
&& \hskip -2cm \int_{t_{\sss \rm e}}^t \!v_e\frac{\bm{n}\!\cdot\!\bm{x}_{\sss\rm 0}}{|\bm{x}_{\sss\rm 0}|^3} \!\ln\!{\frac{|\bm{x}_{\sss\rm 0}|\!+\!\bm{n}\!\cdot\!\bm{x}_{\sss\rm 0}}{|\bm{x}_{\sss \rm e}|\!+\!\bm{n}\!\cdot\!\bm{x}_{\sss \rm e}}} dt \!=\!-\frac{1}{|\bm{x}_{\sss\rm 0}|} \!\ln\!{\frac{|\bm{x}_{\sss\rm 0}|\!+\!\bm{n}\!\cdot\!\bm{x}_{\sss\rm 0}}{|\bm{x}_{\sss \rm e}|\!+\!\bm{n}\!\cdot\!\bm{x}_{\sss \rm e}}} \!-\!\frac{1}{b}\!\Big(\!\arccos\!{\frac{\bm{n}\!\cdot\!\bm{x}_{\sss\rm 0}}{|\bm{x}_{\sss\rm 0}|}} \!-\! \arccos\!{\frac{\bm{n}\!\cdot\!\bm{x}_{\sss \rm e}}{|\bm{x}_{\sss \rm e}|}}\!\Big), \\
&& \hskip -2cm \int_{t_{\sss \rm e}}^t \!\!\frac{v_e}{|\bm{x}_{\sss\rm 0}|^5} \!\ln\!{\frac{|\bm{x}_{\sss\rm 0}|\!+\!\bm{n}\!\cdot\!\bm{x}_{\sss\rm 0}}{|\bm{x}_{\sss \rm e}|\!+\!\bm{n}\!\cdot\!\bm{x}_{\sss \rm e}}} dt\!=\!\frac{1}{3b^2}\!\Big(\!\frac{\bm{n}\!\cdot\!\bm{x}_{\sss\rm 0}}{|\bm{x}_{\sss\rm 0}|^3}\!+\!\frac{2}{b^2}\frac{\bm{n}\!\cdot\!\bm{x}_{\sss\rm 0}}{|\bm{x}_{\sss\rm 0}|}\!\Big)\!\ln\!{\frac{|\bm{x}_{\sss\rm 0}|\!+\!\bm{n}\!\cdot\!\bm{x}_{\sss\rm 0}}{|\bm{x}_{\sss \rm e}|\!+\!\bm{n}\!\cdot\!\bm{x}_{\sss \rm e}}} \!+\!\frac{1}{6b^2}\!\Big(\!\frac{1}{|\bm{x}_{\sss\rm 0}|^2}\!-\!\frac{1}{|\bm{x}_{\sss \rm e}|^2}\!\Big)\!-\! \frac{2}{3b^4}\!\ln\!{\frac{|\bm{x}_{\sss\rm 0}|}{|\bm{x}_{\sss \rm e}|}},\\
&& \hskip -2cm \int_{t_{\sss \rm e}}^t \!\!v_e\frac{\bm{n}\!\cdot\!\bm{x}_{\sss\rm 0}}{|\bm{x}_{\sss\rm 0}|^5} \!\ln\!{\frac{|\bm{x}_{\sss\rm 0}|\!+\!\bm{n}\!\cdot\!\bm{x}_{\sss\rm 0}}{|\bm{x}_{\sss \rm e}|\!+\!\bm{n}\!\cdot\!\bm{x}_{\sss \rm e}}}dt \!=\! \frac{1}{6b^2}\Big(\!\frac{\!\bm{n}\!\cdot\!\bm{x}_{\sss\rm 0}\!}{|\bm{x}_{\sss\rm 0}|^2}\!-\!\frac{\!\bm{n}\!\cdot\!\bm{x}_{\sss \rm e}\!}{|\bm{x}_{\sss \rm e}|^2}\!\Big)\! -\!\frac{1}{3|\bm{x}_{\sss\rm 0}|^3} \!\ln\!{\frac{|\bm{x}_{\sss\rm 0}|\!+\!\bm{n}\!\cdot\!\bm{x}_{\sss\rm 0}}{|\bm{x}_{\sss \rm e}|\!+\!\bm{n}\!\cdot\!\bm{x}_{\sss \rm e}}} \!\nn\\
&& \hskip 2.6cm  -\frac{1}{6b^3}\!\Big(\!\!\arccos\!{\frac{\!\bm{n}\!\cdot\!\bm{x}_{\sss\rm 0}\!}{|\bm{x}_{\sss\rm 0}|}} \!-\! \arccos\!{\frac{\!\bm{n}\!\cdot\!\bm{x}_{\sss \rm e}\!}{|\bm{x}_{\sss \rm e}|}}\!\Big).
\end{eqnarray*}
}
Here, we have made use of $b^2=|\bm{x}_{\sss\rm 0}|^2\!-\!(\bm{n}\!\cdot\!\bm{x}_{\sss\rm 0})^2$.

\section{The photon's velocity in the Kerr-Newman spacetime}\label{Derivation4LightV}
Here we give the derivation for the photon's velocity $\bm{u}$ at the field point $\bm{x}$ with a given direction $\bm{n}$ in the Kerr-Newman spacetime. Since the photon goes along the null geodesic of the background spacetime, we have 
\begin{equation}
g_{\mu\nu}u^{\mu}u^{\nu}=0~,\label{eq:nullGeodecis}
\end{equation}
with $u^{\mu}\equiv (1,\bm{u})$ denoting the 4D velocity of photon. The velocity $\bm{u}$ can be expanded as the sum of the Minkowskian solution and the post-Minkowskian corrections
\begin{equation}
\bm{u}=\bm{u}_{\sss\rm 0PM}+\bm{u}_{\sss\rm 1PM}+\bm{u}_{\sss\rm 2PM}+\cdot\cdot\cdot~,\label{eq:velocity}
\end{equation}
where $\bm{u}_{\sss\rm 0M}$ denotes the Minkowskian solution for the photon's velocity. $\bm{u}_{\sss\rm 1PM}$ and $\bm{u}_{\sss\rm 2PM}$ represent the 1PM and 2PM contributions to the photon's velocity, respectively. When calculating the $k$PM contribution to the velocity, all higher-order contributions should be dropped.

For the zeroth-order approximation, the spacetime is Minkowskian spacetime, the photon's velocity is just
\begin{equation}
\bm{u}_{\sss\rm 0PM}=\bm{n}~,\label{velocity_0PM}
\end{equation}

For the first-order approximation, we keep Eqs.\,(\ref{eq:metric-1nd})-(\ref{eq:metric-3nd}) to the order of $m$, and then substitute them into the null-geodesic equation (\ref{eq:nullGeodecis}), we have
\begin{equation}\label{eq:Nullcondition-1PM}
-1 +\frac{2m}{|\bm{x}|} + \Big(1 +\frac{2m}{|\bm{x}|} \Big)\bm{u}^2 = 0~.
\end{equation}

Substituting Eq.\,(\ref{eq:velocity}) into Eq.\,(\ref{eq:Nullcondition-1PM}), keeping all the terms to the 1PM accuracy only, and then taking into account of Eq.\,(\ref{velocity_0PM}), we can obtain the 1PM contribution to the photon's velocity
\begin{equation}
\bm{u}_{\sss\rm 1PM}=-\frac{2m}{|\bm{x}|}\bm{n}~,\label{velocity_1PM}
\end{equation}

For the second-order approximation, we keep Eqs.\,(\ref{eq:metric-1nd})-(\ref{eq:metric-3nd}) to the order of $m^2$, and then substitute them into the null-geodesic equation (\ref{eq:nullGeodecis}), we have
\begin{equation}\label{eq:Nullcondition-2PM}
\hskip -1cm
-1 +\frac{2m}{|\bm{x}|} -\frac{2m^2\!+\!q^2}{\bm{x}^2} + 2\bm{\zeta}\!\cdot\!\bm{u} + \Big(1 \!+\!\frac{2m}{|\bm{x}|} \!+\! \frac{m^2}{\bm{x}^2}\Big)\bm{u}^2 + \frac{m^2\!-\!q^2}{\bm{x}^4}(\bm{x}\!\cdot\!\bm{u})^2 = 0.
\end{equation}

Substituting Eq.\,(\ref{eq:velocity}) into Eq.\,(\ref{eq:Nullcondition-2PM}), keeping all the terms to the 2PM accuracy only, and then taking into account of Eqs.\,(\ref{velocity_0PM}) and (\ref{velocity_1PM}), we can obtain the 2PM contribution to the photon's velocity
\begin{equation}
\bm{u}_{\sss \rm 2PM}=\bm{n}\Big[\frac{5m^2\!+\!q^2}{2\bm{x}^2}
\!-\!\frac{(m^2\!-\!q^2)(\bm{n}\!\cdot\!\bm{x})^2}{2\bm{x}^4}\!-\!\frac{2 \bm{n}\!\cdot\!(\bm{x}\!\times\!\bm{J})}{|\bm{x}|^3}\Big]~.\label{velocity_2PM}
\end{equation}

Eq.\,(\ref{LightV}) is just the magnitude of the summation of Eqs.\,(\ref{velocity_0PM}),\,(\ref{velocity_1PM}) and (\ref{velocity_2PM}).

\section*{ACKNOWLEDGEMENT}
We would like to thank two referees for their constructive comments and suggestions to improve the quality of this work, and thank Xiaojun Bi for helpful discussions. This work was supported in part by the National Natural Science Foundation of China (Grant Nos. 11647314 and 11747311).

\section*{References}

\bibliographystyle{spphys}       
\bibliography{Reference_20180930}

\end{document}